\documentclass[preprint,appendixfloats]{aastex6}

\def\simlt{\lower.5ex\hbox{$\; \buildrel < \over \sim \;$}}
\def\simgt{\lower.5ex\hbox{$\; \buildrel > \over \sim \;$}}
\def\kms{km~s$^{-1}$}

\def\schi{{\sc Hi}}
\def\schii{{\sc Hii}}
\def\msol{{$M_\odot$}}

\def\sigmahi{{\Sigma_{\rm HI}}}
\def\vlsr{v_{\rm LSR}}

\def\Rsun{R_\odot}
\def\rmkpc{{\rm~kpc}}
\def\fdg{.\!\!^\circ}

\def\tbmax{T_{b,{\rm max}}}

\def\schi{{\sc H$\,$i}}

\newcommand{\disperse}{{DisPerSE}}

\begin{document}


\title{Tracing the Spiral Structure of the Outer Milky Way with Dense Atomic Hydrogen Gas}


\author{Bon-Chul Koo, Geumsook Park, Woong-Tae Kim, and Myung Gyoon Lee} 
\affil{Department of Physics and Astronomy, Seoul National University \\
Seoul 151-747, Korea}

\author{Dana S. Balser}
\affil{ National Radio Astronomy Observatory, \\
520 Edgemont Road, Charlottesville VA 22903-2475, USA}

\and

\author{Trey V. Wenger}
\affil{Astronomy Department, University of Virginia, \\
P.O. Box 400325, Charlottesville, VA, 22904-4325, USA}
\affil{National Radio Astronomy Observatory, \\
520 Edgemont Road, Charlottesville VA 22903-2475, USA}



\begin{abstract}

We present a new face-on map of dense neutral atomic hydrogen (\schi) 
gas in the outer Galaxy. Our map has been produced from  
the Leiden/Argentine/Bonn (LAB) \schi\ 21-cm line all-sky survey
by finding intensity maxima 
along every line of sight  and then by projecting them on the Galactic plane.
The resulting face-on map strikingly reveals the complex spiral structure 
beyond the solar circle, which is characterized by a mixture of 
distinct long arcs of \schi\ concentrations  
and numerous `interarm' features. 
The comparison with more conventional spiral tracers 
confirms the nature of those long arc structures as spiral arms.
Our map shows that the \schi\ spiral structure in the outer Galaxy is 
well described by a four-arm spiral model 
(pitch angle of $12\arcdeg$) with some deviations, and gives a 
new insight into identifying \schi\ features associated with individual arms. 

\end{abstract}




\section{Introduction} \label{sec:intro}

Spiral arms are sites of active star formation with dense interstellar material,
so that young stars, star-forming regions, \schii\ regions, and molecular clouds are 
their conventional tracers. In the outer Galaxy, where star-formation activity is low, 
such `strong' tracers are sparse and neutral atomic hydrogen (\schi) gas is a useful tracer:  
the 21-cm emission line emitted from the transition between 
the two hyperfine levels of the \schi\ atom can be used to trace spiral arms  
that have a higher density of \schi\ gas than the Galactic average.

A conventional approach to explore the \schi\ spiral structure is to 
produce a face-on map of the Galaxy from large-scale  \schi\ 21-cm line surveys. One can transform the observed line flux in a given  
narrow velocity interval along the line of sight 
to the \schi\ column density in the corresponding distance interval.
By doing this for every Galactic longitude ($l$) and 
latitude ($b)$ and by projecting the resulting data cube 
along the direction perpendicular to the Galactic plane,
one can obtain a face-on map of the Galactic \schi\ distribution 
\citep[e.g.,][]{burton1974,binney1998}. 
The main difficulty in this kinetic approach lies  
in transforming the Local-Standard-of-Rest (LSR) velocity ($\vlsr$) to the Galactocentric distance, which is 
usually alleviated by assuming a circular rotation.
Aside from the uncertainty in the adopted Galactic rotation curve,  
the non-circular motions due to streaming along the spiral arms  
and the `velocity crowding' can seriously distort 
the \schi\ structure in real space \citep[e.g.,][]{burton1971}.
This is particularly problematic for the inner Galaxy 
where the line of sight crosses tangential points 
and an ambiguity in distance occurs for a given LSR velocity.  
Harold Weaver, for example, suggested that the 
longitude-velocity $(l,\vlsr)$ 
diagram should be considered as the proper end product \citep{simonson1970}, 
which may be compared with dynamical models.  
For the outer Galaxy, however, these problems are less serious 
except near $l=0\arcdeg$ and $180\arcdeg$ 
where the velocity varies little with heliocentric distance, and    
the \schi\ face-on map can reliably show the spiral structure with some caveats  
if we can separate out dense \schi\ gas associated with spiral arms.

The first \schi\ face-on maps of the outer Galaxy 
appeared in the 1950s \citep{vandehulst1954,westerhout1957,oort1958},
which were soon followed by other maps suggesting different interpretations    
\citep{kerr1969,weaver1970,verschuur1973,weaver1974}.
In order to show the spiral structure, they produced face-on maps either by 
projecting maximum densities in the $z$-direction,
i.e., in the direction perpendicular to the plane,  
on the Galactic plane \citep{westerhout1957,oort1958} 
or by transforming the loci of \schi\ 21-cm line peaks into real space  
and by connecting them \citep{kerr1969, weaver1970,verschuur1973,weaver1974}.  
In his review paper, \cite{simonson1970} 
showed that the two most up-to-date maps at that time, 
i.e., the maps by \cite{kerr1969} and \cite{weaver1970}, 
agree with each other except a few points. 
As we will see in later in \S~2, the sketch by \cite{kerr1969} indeed 
accurately shows the ridges of dense \schi\ concentrations. But without 
demonstrating the association of more conventional spiral tracers, 
the identity of these features as spiral arms could have not been verified.
After these initial works, the efforts to explore the global spiral structure 
of the outer Galaxy using \schi\ face-on maps  
have been rather scarce. As far as we are aware, until 2005, 
the only published \schi\ face-on maps showing the entire outer Galactic plane 
were those in \cite{henderson1982} and \cite{nakanishi2003}. 
Their surface density maps clearly showed three spiral arms; 
the Perseus and Outer (or Norma-Cygnus) arms 
in the northern Galaxy ($l=0\arcdeg$--$180\arcdeg$) 
and the Sagittarius-Carina (hereafter Sgr-Car) arm in the southern Galaxy 
($l=180\arcdeg$--$360\arcdeg$).\footnote{The terms ``southern Galaxy'' and ``northern Galaxy'' are used to 
represent the Galactic areas that are mainly visible from the southern and northern hemispheres, respectively.} 

More recently, as the sensitive \schi\ data of half-degree resolution
from the LAB all-sky survey became available
\citep{kalberla2005,hartmann1997,bajaja2005,arnal2000}, 
new face-on maps of the outer Galaxy 
have been published \citep{levine2006a,nakanishi2016}.  
These new face-on maps of total \schi\ 
surface density showed the above three spiral arm 
structures better than the earlier maps and revealed  
another arm structure in the southern Galaxy. 
But in regions where the Galactic plane is thick and severely warped and flared, i.e.,  
the far side of the northern Galaxy,  
the maps were rather smooth without any hint of an arm-like structure.  
\cite{levine2006a} applied an unsharp masking technique 
which showed the spiral arm features more clearly,  
but it did not reveal new major spiral structures other than 
the ones that can be inferred from the total surface density map. 
They noted that, in several places, the distribution of over-dense regions in their 
perturbed surface density map deviates from the 
spiral model based on \schii\ regions \citep{georgelin1976,wainscoat1992}, 
which has four spiral arms with pitch angles of $12\arcdeg$--$13\arcdeg$. 
Instead, a fit to their perturbed surface density map 
resulted in \schi\ ``spiral arms'' with significantly larger pitch angles of   
21$\arcdeg$--25$\arcdeg$. 

In this paper, we present a new \schi\ face-on map of the outer 
Galaxy and re-examine the \schi\ global spiral structure.
Our map has also been produced from 
the LAB survey data but by employing 
a method similar to those used in early studies;  
we first find local density maxima along sight lines   
and then add up their mass density along the $z$ axis (\S~2).
This face-on map of {\em dense} \schi\ concentrations 
gives a striking view of the complex spiral structure beyond the solar circle,
characterized by a mixture of distinct long arcs  
and numerous interarm features. 
As we will show in \S~3, the known 
conventional spiral tracers are found to be concenrated along the 
\schi\ ridges, which verifies the identity of 
long arc structures as spiral arms. 
We will further show that the global \schi\ spiral structure in the outer Galaxy 
connects smoothly to the massive star-forming regions 
in the inner Galaxy and that they are described well  
by a four-arm spiral model (pitch angle $12\arcdeg$) 
with some deviations. 
We will also discuss \schi\ features 
associated with individual arms in \S~3. 

\section{Data and Face-on Map Production}
\label{sec:data}

\begin{figure}
\centering
\vspace*{2.0truecm}
\includegraphics[width=5.0in]{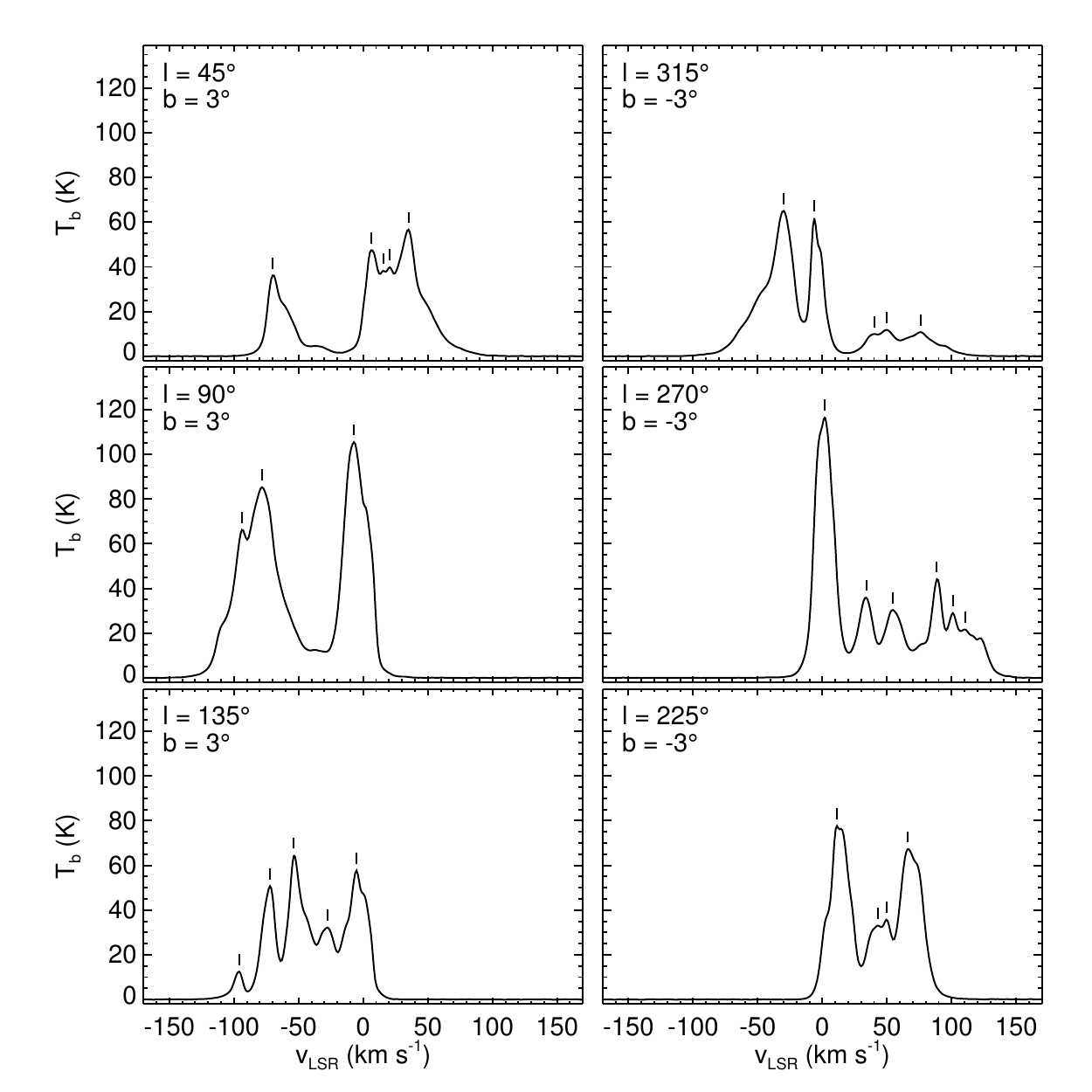}
\caption{ 
Sample \schi\ spectra of the LAB survey with the local maxima found by the code 
PEAKFINDER marked (see text for more details).  
The portions of the profiles corresponding to the emission from the \schi\ gas outside 
the solar circle for $l=0\arcdeg$--$180\arcdeg$ ($180\arcdeg$--$360\arcdeg$) 
are those at $\vlsr< 0$  ($> 0$).
\label{fig:fig1}}
\end{figure}

We use the LAB \schi\ 21-cm line survey data.
The survey combined three independent surveys of angular resolution of 30$'$--36$'$
to produce all-sky data at $0\fdg5$ pixels 
with an rms noise of 0.07--0.09~K in brightness temperature.
We first determine the local maxima in each line profile.
We use the IDL procedure PEAKFINDER\footnote[2]{The code can be obtained at 
\url{http://132.248.1.102/~morisset/idl_cours/IDL/index_local.htm}.},
developed by Christophe Morisset (UNAM, Mexico). The code finds local maxima from their 
derivatives and then determines the significance of a peak from its width 
and weight. 
The width of a peak is the number of data points that have positive derivative at its 
left plus the number of data points with negative derivative at its right, and  
the weight is the sum of  the absolute values of the derivatives inside the width.
We adopted the peaks with the weights $\ge 0.5$ and $T_b > 0.5$~K, 
which adequately identifies 
all distinct intensity maxima in the spectra (Figure \ref{fig:fig1}).

The $(l,b,\vlsr)$ cube of the local maxima brightness temperature, 
$T_{b,\rm max}(l,b,\vlsr)$, is used to fill up the three-dimensional spatial data cube of 
the \schi\ mass density $\rho_{\rm H}(x,y,z)$. 
We first determine $(l,b,\vlsr)$ of each $(x,y,z)$ pixel and 
obtain its $T_b$ from a trilinear interpolation
using the IDL routine INTERPOLATE.
The conversion from $(x,y,z)$ to $\vlsr$ is carried out   
by using a flat rotation curve
with  the distance to the Galactic center $R_0=8.34$~kpc   
and the rotational speed of the Sun $\Theta_0$ = 240~\kms, which has been suggested  
from the parallax-based distances and proper motions of high 
massive star forming regions (HMSFRs) \citep{reid2014,reid2016b}.
The Sun's rotation speed is considerably greater than the IAU 
value (220~\kms), and the authors attributed the difference to  
the biased estimate of $\Theta_0$ from the \schi\ data 
due to the curvature in the rotation curve \citep{reid2016b}.
From $T_b(x,y,z)$, the H column density per unit velocity $\Delta  N_{\rm H}$  
is obtained at every voxel and it is converted to  
the \schi\ mass density $\rho_{\rm H}(x,y,z)$ dividing by $|dr/dv_{\rm LSR}|$, i.e.,
\begin{equation}
{\rho_{\rm H} (x,y,z) \over m_{\rm H}}= {\Delta N_{\rm H}(x,y,z) \over  |dr/dv_{\rm LSR}| } 
=0.591 { T_b(x,y,z) \over |dr/dv_{\rm LSR}| } 
{\tau_\nu \over 1 - e^{-\tau_\nu}}~~~{\rm cm}^{-3}, 
\end{equation}
where in the second equation, 
$\tau_\nu$ is the 21-cm optical depth and $T_b$ and  
$|dr/dv_{\rm LSR}|$ are in units of K and pc (km s$^{-1}$)$^{-1}$, respectively.
The optical depth has been estimated   
by assuming a constant spin temperature 155~K 
\cite[see, for example,][for details]{levine2006b}.
We emphasize that, since we sample only $T_b$ 
at intensity maxima in the spectra, the resulting data cube $\rho_{\rm H}(x,y,z)$ 
represents a very limited mass/volume of dense \schi\ gas, 
and the column density projected to the Galactic plane will be much less than 
the total \schi\ column density of the Galactic disk. 
But as far as the intensity peaks are due to  
\schi\ concentrations, the resulting face-on map should be a 
representative map of dense \schi\ gas. 

Our spatial data cube covers  
an area of $50\times 50$~kpc$^2$ in the plane 
and $\pm 6$~kpc in the direction perpendicular to the plane.
This $z$-range covers most of the warped Galactic plane:  
the height of the northern Galactic plane above the $b=0\arcdeg$ plane 
increases almost linearly with Galactocentric distance 
and reaches $+4$~kpc at $R=22$~kpc at $l=90\arcdeg$,  
while that of the southern Galactic plane increases with $R$, 
reaching $1$~kpc (at $l=270\arcdeg$) below the $b=0\arcdeg$ plane at $16$~kpc,  
and then decreases back to the $b=0\arcdeg$ plane \citep{burton1988,levine2006b}.
The voxel size of the spatial cube is 50~pc. 
The angular resolution ($30'$) of the LAB survey corresponds to  
42~pc at 5~kpc and 170~pc at a distance of 20~kpc, so that 
the cube undersamples the area near the Sun 
but oversamples the area on the opposite side of the Galactic center 
\citep[see][]{levine2006a}. 
The two-dimensional map of the \schi\ mass 
surface density ($\sigmahi(x,y)$) in this paper is obtained by summing the above  
three-dimensional cube along the direction perpendicular to the plane
from $-3$ to $+6$~kpc and then  
smoothing by a Gaussian kernel with FWHM of 500~pc.

The adopted rotation curve has a small effect on the resulting face-on map. 
If we use the flat rotation 
curve with $\Rsun=8.5$~kpc and $\Theta_0 = 220$~\kms,  
all the \schi\ features slightly shift away from the Galactic center but 
their morphology remains the same. 
On the other hand, if we use the rotation curve recently proposed 
to correct the large \schi\ 
surface density contrast across the Sun-Galactic center line \citep{levine2006b}, 
the curvature of \schi\ features slightly changes near galactic longitudes $l=0\arcdeg$ and $180\arcdeg$, 
where an epicyclic streaming motion correction has been made. Since we will be comparing the \schi\ spiral structures to 
the distribution of HMSFRs in the inner Galaxy, we use 
the rotation curve based on the HMSFR data \citep{reid2014,reid2016b}.

\section{Results and Discussion}
\subsection{\schi\ Face-on Map of the Outer Galaxy}
\label{sec:res1}

Figure~\ref{fig:fig2} is our \schi\ face-on map of the outer Galaxy, which strikingly    
shows the complex spiral structure of the \schi\ concentrations in the outer Galaxy. We see 
several prominent long ($\simgt 20$~kpc) arcs  
that might be segments of major spiral arms. 
There are also numerous ridges either connecting these  
spiral arm features or branching out from them. 
In Figure~\ref{fig:fig3} (a)-(b), we compare our map with  
the sketches of \schi\ spiral structure by 
\cite{kerr1969} and \cite{weaver1970} \citep{simonson1970}.
It is clear that these sketches of early \schi\ studies 
properly showed the locations and shapes of the major features.
Some of the Kerr's ridges are slightly shifted from the bright 
features in our map, which might be due to the 
different data sets used in the two studies.
We also compare our map with the face-on map of perturbed surface densities obtained by \cite{levine2006a} in Figure~\ref{fig:fig3} (c)-(d).
Note that the Levine's map was obtained by processing the 
total surface density map so that it shows the areas of excess   
{\em surface densities}, while our map is obtained by processing the 
data cube {\em before} integrating along the $z$-axis and  
it shows dense regions traced by intensity peaks.
The appearances of the two maps are quite different, although the locations of 
prominent features agree in general. 
Our map shows the continuous arm structures 
more clearly in great detail and also reveals new features. 
In the far side of the northern Galaxy where 
the Galactic plane is severely warped and flared, 
for example, there is a distinct thin and long arc structure at $(l,R)=(15\arcdeg,12~{\rm kpc})$
to $(70\arcdeg, 16~{\rm kpc})$ in our map, 
but no coherent arm feature is apparent in the map of \cite{levine2006a}. 
This structure is  a segment of the outer Scutum-Centaurus 
(hereafter Sct-Cen) arm \citep[][see below]{dame2011}, and it  
demonstrates the advantage of our technique in tracing the spiral structure.

The identity of the \schi\ arc structures in Figure~\ref{fig:fig2} 
as spiral arms can be confirmed  
by comparing to the distribution of other more conventional spiral tracers.
In Figure~\ref{fig:fig4} (top panel), we overlay the distribution of \schii\ regions and CO clouds
on the \schi\ face-on map. 
The distances to these sources are derived in the 
same way as the \schi\ concentrations 
in order to avoid an offset due to the uncertainties in converting 
observed LSR velocities to Galactocentric distances. 
The red crosses represent the \schii\ regions in the outer Galaxy from the 
catalog of the all-sky {\em Wide-Field Infrared Survey Explorer} 
({\em WISE}) survey \citep{anderson2014}, while 
the other symbols mark CO molecular clouds detected 
in the outer Galaxy \citep{may1997,vazquez2008,dame2011, sun2015,du2016}.  
Only the area between $l=100 \arcdeg$ and $150\arcdeg$ has been 
fully mapped in CO emission at sufficient 
angular resolution ($\sim 1'$) \citep{sun2015,du2016}.
We will discuss the association of these sources 
with individual \schi\ spiral-arm features in the following section, but their 
association with the \schi\ filamentary structures is clear in Figure~\ref{fig:fig4}.  
For example, we see strong concentration of \schii\ regions 
along the \schi\ arc structure at $(285\arcdeg, 8.3~\rmkpc)$ to $(320\arcdeg,10~\rmkpc)$ 
which is the well-known Sgr-Car arm, while  
the clustered \schii\ regions between $l=105\arcdeg$ and 
$155\arcdeg$ at $R=10$--12 kpc represent a segment of the Perseus arm.
The concentrated \schii\ regions along the solar circle near the Sun 
belong to the Local arm which has been known as a minor spiral arm 
but with star formation rate comparable to those of major spiral arms \citep{xu2016}. 
The \schi\ structure associated with the Local arm is not particularly apparent in our map.
Figure~\ref{fig:fig4} also shows that some interarm \schi\ structures have associated 
\schii\ regions indicating massive star formation in these structures, e.g., 
see the $\sim 4$ kpc-long ridge 
from $(40\arcdeg, 14~{\rm kpc})$ to $(48\arcdeg, 12~{\rm kpc})$.
Similar figures comparing the distribution of 
spiral-arm tracers with the \schi\ map of \cite{nakanishi2003} or 
that of \cite{levine2006a} may be found in other works 
\citep{efremov2011,hou2014}.
We should also mention that the association of \schi\ structures with 
CO clouds has been explored in the above mentioned 
CO works \citep{dame2011,sun2015,du2016}.

There are caveats for Figure~\ref{fig:fig2}. 
First, some spiral arm segments 
could have been shifted and/or deformed by non-circular motions.
Spiral arms with strength similar to those in the Milky way are known 
to excite non-circular streaming motions of order 10--20~\kms\  
\citep[e.g.,][]{fresneau2005,kim2014,errozferrer2015}, 
which can shift their locations in real space by a few kpc during the mapping.
It is difficult to locate such shifts or deformations in Figure~\ref{fig:fig2} 
without prior information. 
We mention some known features in the next section. 
Second, there are artifacts around the solar circle ($R=8.34$ kpc) due to  
the \schi\ gas in the solar neighborhood, i.e., 
the local \schi\ gas at $\vlsr\ne 0$~\kms\  
is mapped to the other side of the solar circle in the real space.
Such artifacts can be identified by looking at Figure~\ref{fig:fig4} (bottom frame), 
which shows the mass-weighted mean 
height, $\langle z \rangle \equiv \int z \rho(z)dz/\sigmahi$,
of the \schi\ concentrations. 
We can see that, for example, the strong emission 
from $l\approx 300\arcdeg$ to $345\arcdeg$ just outside 
the solar circle 
has a very large scale height, which indicates that it is mostly
due to local gas having a small positive $\vlsr$.  
Third, the surface density in Figure~\ref{fig:fig2} is a  
fraction of total \schi\ surface density.
It represents the projected surface density of 
dense \schi\ gas {\em at intensity peaks},   
averaged over $\sim 500$ pc (see \S~2).
The comparison to the total surface density 
map, which may be found in \cite{levine2006a} or 
\cite{nakanishi2016} and is not shown here,
indicates that the surface density in 
Figure \ref{fig:fig2} is $\simlt 10$\% 
of the total surface density. 
Before the smoothing, the surface density within a pixel (50 pc)  
is as large as 30\% of the total surface density. 
Figure~\ref{fig:fig2} shows the relative brightnesses  
of \schi\ structures, but it should not be used for a quantitative 
comparison with other spiral tracers, e.g., CO intensity. 

\subsection{\schi\ Spiral Arms and the Four-arm Spiral Model}
\label{sec:res2}






Figure~\ref{fig:fig2} makes it clear that  the \schi\ concentrations in the outer 
Galaxy have complex spiral features but with enough regularity suggesting 
a global spiral pattern. We compare the \schi\ distribution 
to the four-arm spiral model with a constant pitch angle along each arm 
\citep[see][]{georgelin1976,wainscoat1992,churchwell2009, efremov2011, vallee2014, vallee2015,bobylev2014,hou2014}. 
In this model, the spiral pattern has four, roughly equally-spaced, major arms, i.e., 
the Sgr-Car, Perseus, Outer, and Sct-Cen arms, 
with a mean global pitch angle between 12 and 14 deg.\footnote{The spiral arms are referred to by different names, e.g., the Norma-Cygnus arm for the Outer arm, and the Scutum-Crux arm for the 
Scutum-Centaurus arm \citep{vallee2014}.}
It has been shown that some \schi\ 
arc structures are  well described by this global model 
\citep[][and references therein]{levine2006a,nakanishi2016}.
On the other hand, observations of external galaxies 
indicate that the pitch angle can vary considerably along 
individual arms \citep[][and references therein]{savchenko2013,davis2014,honig2015}. 
In the Milky Way, trigonometric parallax measurements of 
HMSFRs also found a considerable variation of  
pitch angles ($7\arcdeg$--$20\arcdeg$) among 5--10 kpc segments of spiral arms 
\citep[][and references therein]{reid2014,reid2016a}.
Our \schi\ face-on map also shows such variations of pitch 
angles along some long ridges, 
but here we will be focusing on the global spiral structure. 
In Figure~\ref{fig:fig5} (top frame), 
we plot the distribution of the HMSFR with accurate distances \citep{reid2016a} 
on our \schi\ map.  
Recall that the kinematic distances to the \schi\ concentrations have been  
computed by using the rotation curve obtained from the HMSFR data \citep{reid2014}, 
so that it is justified to compare these two different category sources, i.e., 
\schi\ concentrations and HMSFRs.   
We find that the two well-established \schi\ spiral arm features, i.e.,  
the Sgr-Car arm in the southern Galaxy and the outer 
Sct-Cen arm in the northern Galaxy (see below), 
can be well connected to the HMSFRs in the inner Galaxy 
by logarithmic spirals over 180$\arcdeg$ and 
360$\arcdeg$ in Galactocentric azimuth, respectively.
For the other two arms, i.e., the Perseus and the Outer arms, there is  
currently considerable confusion in the identification of their associated \schi\ features 
as we will discuss below, 
but  in our map we can clearly identify \schi\ long arcs 
matching these arms in the four-arm spiral model. 

In order to facilitate discussion and also for quantitative analysis, 
we need traces of \schi\ ridges giving the positions of \schi\ arm structures. 
This has been done in an $(l,\vlsr)$ diagram where we can 
see the structures even around $l=0\arcdeg$ and 180$\arcdeg$,
and the procedure is explained in the Appendix.
Here we give a brief summary.  
We first obtained a two-dimensional $(l,\vlsr)$ map 
by integrating  the three dimensional cube of local peak intensities, $\tbmax(l,b,\vlsr)$, 
along the $b$ axis and determined the traces of all \schi\ ridges 
by delineating the \schi\ peak positions in an unbiased way.  
We then choose the ridges/traces that are thought to be 
the segments of major  arms. For the well-established arm features or for the  
coherent long arc structures, this procedure is rather straightforward.
But in the areas with complicated structures,   
e.g., between $l=100\arcdeg$ and 150$\arcdeg$, 
or for the features of confusing identity (see below),
this procedure is subjective. 
It is, however, clear that two- or 
three-arm models cannot describe the 
regularity of the observed \schi\ spiral structure, and 
the selected ridges are allocated to one of the 
four spiral arms.
The locations of adopted ridges in the $(l,\vlsr)$ diagram and 
their arm identification are summarized in Table A1. 
Similar spiral-arm traces  
had been obtained by \cite{weaver1974} and \cite{reid2016a}, and 
we compare ours to those of \cite{reid2016a} in the Appendix. 
They agree with each other for most arm features, but some ridges 
are assigned to different spiral arms (see below).
The selected spiral arm traces in the $(l,\vlsr)$ diagram 
are then mapped to the real space, which are 
shown as dotted lines in Figure~\ref{fig:fig5} (bottom frame).
Note that we selected two ridges for some portions of spiral arms
when there are two parallel branches 
or when the arms bifurcate at their ends.
We emphasize that these traces are mainly to facilitate the discussion.
In the quantitative analysis below, 
only the well-established traces of the Sgr-Car and the Sct-Cen arms 
are used.

We can now derive the parameters 
of the Sgr-Car and Sct-Cen spiral arms that have  
well-established \schi\ features. 
As we mentioned in \S~\ref{sec:res1}, 
the thin and long arc structure at $(l,R)=(15\arcdeg,12~{\rm kpc})$
to $(70\arcdeg, 16~{\rm kpc})$ in Figure~\ref{fig:fig5} 
is the segment 
of the outer Sct-Cen arm \citep{dame2011}
and, together with the HMSFRs in the inner Galaxy, 
it delineates the Sct-Cen arm over 360$\arcdeg$ 
in Galactocentric azimuth $\phi$. 
Figure~\ref{fig:fig6} is a plot of   
$\log R$ versus $\phi$ for all spiral tracers,   
clearly showing that the Sct-Cen arm can be modeled as 
a logarithmic spiral with a well-defined pitch angle.
We fit the distribution of the HMSFRs and the \schi\ trace 
by $\log(R(\phi)/R_0)=(\phi-\phi_0) \tan\psi$, 
where $\phi=\phi_0$ at $R=R_0$ and 
$\psi$ is the arm pitch angle, that is, the angle between 
the tangents of the spiral arm and the circle at that point.
The weights are given to the \schi\ traces so that  
their total weight is equal to that of the HMSFRs, but the result does not 
depend on the details of the weighting.
The least-squares fit to $\log R$ versus $\phi$ 
yields $\psi=12\fdg4\pm 1\fdg8$
and $\phi_0=288\arcdeg\pm 47\arcdeg$.
The derived pitch angle is 
consistent with the mean global pitch angle obtained in previous studies, i.e., 
$\sim 12\arcdeg$ \citep{vallee2014}. 
The Sgr-Car arm also has a well-known \schi\ 
feature which extends from $(285\arcdeg, 8.3~\rmkpc)$ 
to $(320\arcdeg,10~\rmkpc)$ in Figure~\ref{fig:fig5}. 
If we do the same analysis for the Sgr-Car arm, we
obtain  $\psi=10\fdg9\pm 2\fdg8$
and $\phi_0=225\arcdeg\pm 58\arcdeg$. 
But considering the small ($\sim 180\arcdeg$) coverage in $\phi$
and the large error bar, we may instead adopt the same pitch angle ($12\fdg4$) 
assuming symmetric arms and fit only $\phi_0$. 
In this case,  we obtain $\phi_0=$ $223\arcdeg\pm 17\arcdeg$, and   
the spacing between the Sgr-Car and Sct-Cen arms at the solar circle is 9.5 kpc. 
For the other two arms,  i.e., the Perseus and Outer arms, 
their identification is controversial (see below). 
So we just draw spirals in Figure~\ref{fig:fig5} obtained by 
rotating the above two spirals by $180\arcdeg$ in $\phi$, i.e.,  
$\phi_0=108\arcdeg$ and $43\arcdeg$ with the same pitch angle 
($12\fdg4$) for the 
Perseus and Outer arms, respectively, to facilitate the discussion.
The parameters of the spiral arms in Figure~\ref{fig:fig5} are summarized in
Table~\ref{table:tbl1}. In the following we briefly discuss individual 
major arm features:


\bigskip\noindent
(1) Sgr-Car arm

The Sgr-Car arm appears as a coherent, narrow structure 
extending over 20 kpc from $l=285\arcdeg$ to $345\arcdeg$. 
This \schi\ structure as well as its association with \schii\ regions
has been well established \citep[e.g.,][]{georgelin1976,hou2014}. 
Figure~\ref{fig:fig5} shows that 
there are several faint interarm ridges connecting to the Perseus arm  
between $l=305\arcdeg$ and $330\arcdeg$. 
The long ridge from $(318\arcdeg, 12~{\rm kpc})$ to $(330\arcdeg, 16~{\rm kpc})$ 
is noticeable.
The \schi\ structure connects well to the HMSFRs in the inner Galaxy by 
a logarithmic spiral with a pitch angle of $12\arcdeg.4$, 
considerably greater than $6\fdg9$ 
obtained from the HMSFRs in the inner Galaxy alone \citep{reid2014}.

\bigskip\noindent
(2) Perseus arm

In the northern Galaxy, 
the Perseus arm in \schi\ appears 
to be composed of several short segments 
between $l=60\arcdeg$ and 155$\arcdeg$. 
This section of the Perseus 
arm has been well known \citep{georgelin1976}, but 
its location and shape are uncertain due to non-circular motions. 
At $l=110\arcdeg$--155$\arcdeg$, 
the Perseus arm sources have peculiar velocities 
of 10--20~\kms, so that 
their locations are considerably ($\sim 2$~kpc) shifted outwards 
\citep{xu2006,reid2014}.  
This can be seen in Figure~\ref{fig:fig5} 
where the locations of HMSFRs 
expected from their LSR velocities are marked by thin soild lines
stretching out from their symbols.  
There is an \schi\ arc structure that extends from 
$l=90\arcdeg$--$110\arcdeg$ at $R=11.5$ kpc 
and is overlapped with many \schii\ regions. 
The structure has $\vlsr=-70$~\kms\ to $-60$~\kms\
and could be an interarm structure (see the Appendix).
\cite{zhang2013} and \cite{reid2016a} pointed out a dearth of 
HMSFRs in the Perseus arm between $l=50\arcdeg$ and  
$80\arcdeg$. Figure~\ref{fig:fig4} 
also shows a clumpy distribution of {\it WISE} \schii\ regions with gaps in this longitude range 
\citep[see also Figure 5 of][]{reid2016a}. 
The \schi\ concentrations fill in these gaps and provide complementary 
information for  tracing spiral arms.
The model spiral arm opposite to the Sct-Cen arm 
with the same ($12\fdg4$) pitch angle  can describe
this section of the Perseus arm reasonably well (Figure~\ref{fig:fig5}). 

In the southern Galaxy, the identification of the Perseus arm is controversial. 
We note that in several previous studies  
the outermost \schi\ spiral feature (that we identify as the Outer arm) 
in Figure~\ref{fig:fig5} has been identified  
as the Perseus arm \citep{nakanishi2003,levine2006a,reid2016a}. 
We, however, identify the relatively diffuse \schi\ ridges  
running along the inside of the 
model Perseus arm as the Perseus arm trace.
The trace is well-defined between $l\approx 200\arcdeg$ and $240\arcdeg$,
while, between $l\approx 240\arcdeg$ and $300\arcdeg$,
there appears to be two parallel ridges  
that can correspond to the Perseus arm trace  (see also \S~\ref{sec:res3}).
And at its ends ($l\approx 300\arcdeg$--$330\arcdeg$), 
the \schi\ arm appears to bifurcate.   
As we see in Figures~\ref{fig:fig4} and \ref{fig:fig5},  
there are HMSFRs around $l=230$--$240\arcdeg$ and    
many CO clouds between $l\approx 205\arcdeg$ and $270\arcdeg$ 
\citep{vazquez2008} along our Perseus arm trace, which 
supports its identity as a spiral arm trace.  
We can make the model arm to match the \schi\ feature better  
by increasing $\phi_0$ slightly, i.e., 
$\phi_0=117\arcdeg$ instead of $108\arcdeg$. 
\cite{levine2006a} instead fit this segment by a new logarithmic 
spiral arm of large (25$\arcdeg$) pitch angle, while 
\cite{nakanishi2016} identified it as an extension of the Local arm. 
\cite{mgriffiths2004}, however, identified the ridge between 
$=253\arcdeg$ and $321\arcdeg$ as a separate arm (the ``Distant Arm'') that can be 
smoothly connected to the Outer arm by a small ($9\arcdeg$) pitch angle. 
It is worth to note that there are essentially no {\it WISE} \schii\ regions 
at $l\simgt 270\arcdeg$ associated with the \schi\ structure, 
but only {\it WISE} HII regions with measured
velocities are shown and therefore the lack of \schi\ regions may be due
to less sensitive line surveys of the Southern Galaxy \citep{brown2017}.

\bigskip\noindent
(3) Outer arm

The Outer arm appears as the most prominent 
\schi\ feature in the northern Galaxy.
The segment between 
$(20\arcdeg, 8.3~{\rm kpc})$ 
and $(80\arcdeg,13~{\rm kpc})$ is the brightest feature and 
has been well known \citep{henderson1982,nakanishi2003}. 
But it is systematically shifted outwards from the 
model spiral arm by $\simlt 2$ kpc, which could be 
due to non-circular motions associated 
with streaming along the spiral arm. 
Between $l=80\arcdeg$ and $110\arcdeg$, 
the arm appears to split into two segments, and at larger 
$l$ ($110\arcdeg$--$150\arcdeg$) 
the structure becomes complex 
with ridges blending each other. 
\cite{du2016} detected CO clouds 
in this area ($l=100\arcdeg$-- $150\arcdeg$) and 
investigated the associated \schi\ structure.

In the southern Galaxy, 
we identify the outermost, narrow and bright coherent feature between 
$l=210\arcdeg$ and $300\arcdeg$ as the Outer arm, which appears 
as a smooth coherent structure running 
closely parallel to the model Outer arm. 
The reason that we associate this structure to the Outer arm 
and not to the Perseus arm is mainly because there is another
distinct arm structure that can be assigned to the Perseus arm 
as we discuss above. 
Early \schi\ studies also detected 
the bright ridges of this structure and identified them as an outer spiral arm beyond 
the Perseus arm \citep{kerr1969,davies1972} (Figure \ref{fig:fig3}).
In the face-on map, however, there is a large positional  
jump in the Outer arm trace from the northern  
to the southern Galaxy across the gap around $l=180\arcdeg$ (Figure \ref{fig:fig5}), which 
led people to identify the bright portion between $l=210\arcdeg$ and $260\arcdeg$ 
as the Perseus arm in some previous studies 
\citep{nakanishi2003,levine2006a}.
We will address this issue in the next section, but 
basically there is no jump in $(l,\vlsr)$ space, and 
the jump in the face-on map was generated 
in transforming $\vlsr$ to  the Galactocentric distance. 
On the other hand, 
the bright portion is relatively straight and is systematically shifted inwards from the 
logarithmic spiral by $\simlt 2$~kpc.
\cite{efremov2011} pointed out that its morphology is  
similar to the so-called ``kneed'' spiral arms in external galaxies.
But it could be also due to streaming motions along 
the spiral arm.   
\cite{vazquez2008} detected CO clouds associated with the 
\schi\ ridge between $l=190\arcdeg$ and $255\arcdeg$, and 
pointed out that their distribution and the \schi\ structure 
are well matched to the Outer arm 
in the four-arm spiral model. 
At its ends, between $l=285\arcdeg$ and $300\arcdeg$, 
the Outer arm also appears to bifurcate.

\bigskip\noindent
(4) Sct-Cen arm

The outer Sct-Cen arm appears as 
a distinct, $\simgt 20$~kpc-long \schi\ arc structure in the outermost northern Galaxy 
in our face on map. The bright segment between  
$(15\arcdeg, 12~{\rm kpc})$ and $(70\arcdeg, 16~{\rm kpc})$ 
had been identified in early \schi\ studies 
\citep{kerr1969,weaver1974} and was rediscovered recently by \cite{dame2011}.
There is a prominent, $\sim 4$ kpc-long \schi\ ridge 
between $(38\arcdeg, 13~\rmkpc)$ and $(46\arcdeg,12~\rmkpc)$ with many \schii\
regions, which appears to be an interarm structure (see the Appendix).
At $l\sim 70\arcdeg$, the arm structure becomes faint and bends inwards, and 
at $l\sim110\arcdeg$, it appears to blend with other arc structures. 
\cite{sun2015} claimed the `flared' feature between $l=110\arcdeg$ and $150\arcdeg$ 
as the continuation of the Sct-Cen arm based on their detection of CO clouds.

In the southern Galaxy, there is strong \schi\ emission  
along the Sct-Cen arm at $l\approx 320\arcdeg$--$345\arcdeg$ 
just outside the solar circle (Figure~\ref{fig:fig4}). As we mentioned in \S~3.1, most of this emission is 
from the local \schi\ gas in the solar neighborhood with slightly positive 
$\vlsr$. But we find that, if we generate a surface density map of \schi\ gas 
near the Galactic plane, e.g., within $|z| \simlt 0.5 {\rm kpc}$ from the $b=0\arcdeg$ plane,  
we can see an \schi\ feature that corresponds to the Sct-Cen arm. 

\subsection{The Perseus and Outer Arm Issue}
\label{sec:res3}


As mentioned in section~\ref{sec:res2}, 
there is considerable confusion in the extension of the Perseus and 
Outer arms in the southern Galaxy ($l=180\arcdeg$--360$\arcdeg$).
A major issue is the identification of the outermost, bright \schi\ ridge between 
$l=210\arcdeg$ and $280\arcdeg$: {\em Is it an extension of the 
Outer arm or the Perseus arm?}
To summarize, early \schi\ studies \cite[e.g.,][]{kerr1969,davies1972} identified it 
as an outer spiral arm beyond the Perseus arm. 
We also identified it as an Outer arm extension because 
there is another arm feature that can be assigned to the Perseus arm. 
In contrast, several previous studies identified it 
as an extension of the Perseus arm that can be connected smoothly across 
$l=180\arcdeg$ with a large ($16\arcdeg$--$24\arcdeg$) pitch angle 
\citep{nakanishi2003,levine2006a,reid2016a}. 

The confusion is partly due to the difficulty in tracing the spiral structure near 
$l=180\arcdeg$ in the face-on map where one cannot obtain reliable kinematic distances.
So it is helpful to look at an $(l,\vlsr)$ map. 
Figure \ref{fig:fig7} is an $(l,\vlsr)$ map of dense \schi\ concentrations 
in the 2nd ($l=90\arcdeg$--180$\arcdeg$) and 3rd  
($l=180\arcdeg$--270$\arcdeg$) Galactic quadrants (hereafter Q2 and 
Q3, respectively) produced 
by integrating $T_b(l,b,\vlsr)$ in \S~\ref{sec:data} from 
$b=-5\arcdeg$ to $+5\arcdeg$ (see Appendix for other maps integrated over a wider latitude range).  
The ridges that we associate with 
the Perseus and Outer arms are marked in green and cyan colors, respectively 
(see Figure~\ref{fig:fig5} for their locations in real space).
Figure~\ref{fig:fig7} clearly shows that the Perseus and 
Outer arm ridges are both prominent in Q3, e.g., the Perseus arm ridge between
$(l,\vlsr)\approx (190\arcdeg, 20~{\rm km~s}^{-1})$ and $(240\arcdeg, 70~{\rm km~s}^{-1})$ 
and the Outer arm ridge between 
$(l,\vlsr)\approx (190\arcdeg, 20~{\rm km~s}^{-1})$ 
and $(260\arcdeg, 110~{\rm km~s}^{-1})$.
At larger $l$, each of them appears to split into two and the structure becomes complicated.
The CO clouds and \schii\ regions along the two ridges support 
the nature of these ridges as spiral arms, and the location 
of the Perseus-arm HMSFRs with accurate distances \citep{reid2016a}  
around $l=230\arcdeg$--$240\arcdeg$
further supports their identity as the Perseus and Outer arm ridges.
 
Connecting the Perseus and Outer arm ridges in Q3 to those 
in Q2 across $l=180\arcdeg$ is not straightforward.
First, the two ridges merge with each other at $l\simlt 200\arcdeg$. 
Indeed, between $l=190\arcdeg$ and $180\arcdeg$, 
all \schi\ features merge into one single ridge because 
of small $|d\vlsr/dr|$.
Then in Q2, the \schi\ spiral structure is  
complicated. We have traced prominent \schi\ ridges to 
yield a coherent structure and again the 
strong spiral tracers along these ridges support their spiral arm nature, but  
individual segments show considerable deviations from a smooth pattern.
The positional jump in the Outer arm trace across 
the gap near $l=180\arcdeg$ in the face-on map, therefore, 
cannot be strong evidence against 
its identity as a coherent spiral structure. It is also interesting to note that the Outer arm \schi\ ridges are generally faint in Q2 but bright in Q3, 
which appears to be implausible considering that 
the Outer arm in Q3 is at a larger Galactic radius. 
We think that this could be because our face-on map shows dense 
\schi\ concentrations, while the outer Galactic disk in Q2 is 
severely warped and flared so that \schi\ gas there is relatively diffuse. 
Indeed the contrast in \schii\ regions and CO clouds 
between Q2 and Q3 for the Outer arm is similar. 
 

\section{Summary}
\label{sec:disc}

\begin{figure}[b]
\centering
\includegraphics[width=7.0in, trim={0 1cm 0 0}]{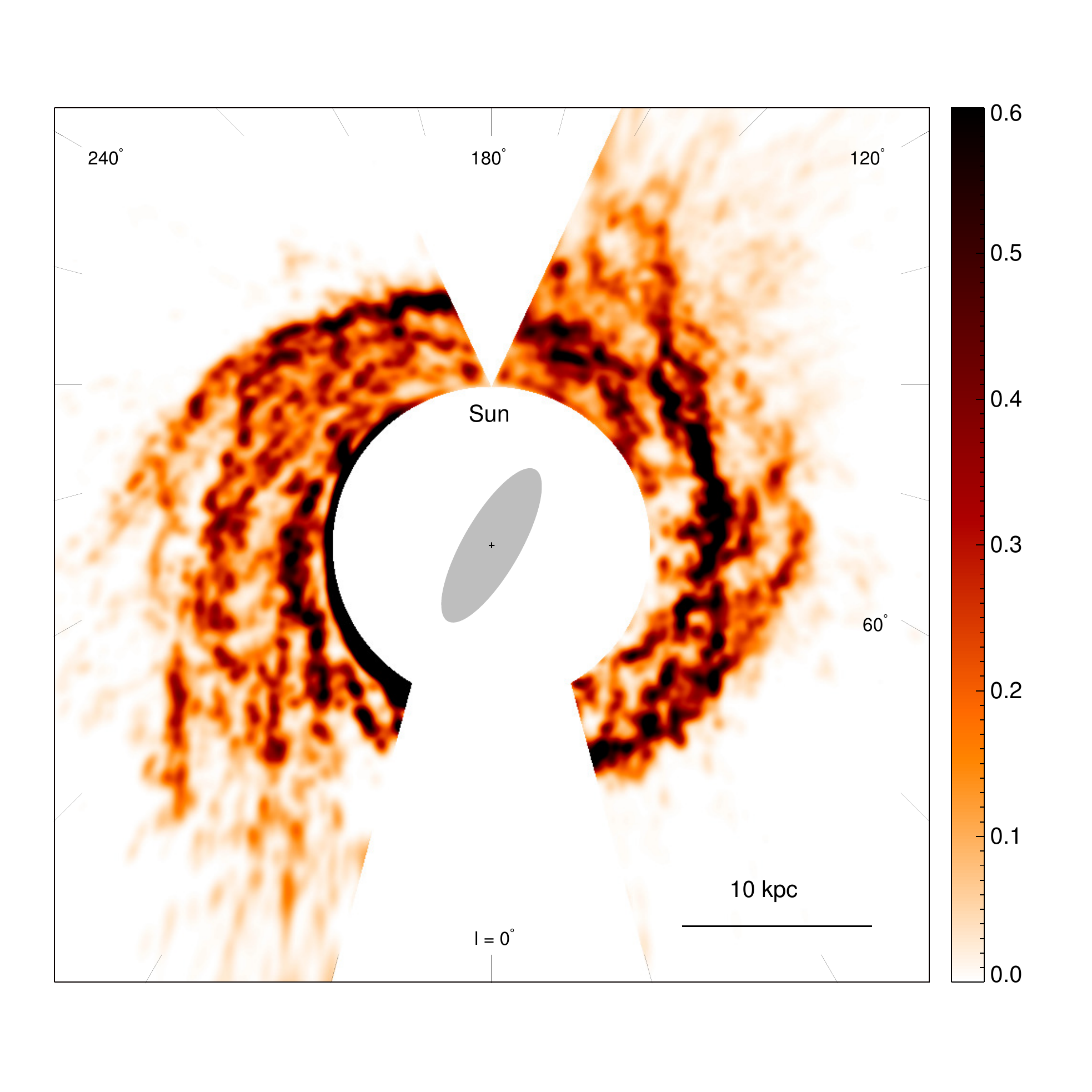}
\caption{
Face-on map of \schi\ concentrations in the Galaxy beyond the solar circle
obtained in this paper. 
The color scale represents the mass surface density ($\sigmahi$)
of \schi\ concentrations in units of $M_{\odot}$~pc$^{-2}$. 
It should be noted that the locations of \schi\ concentrations could have been 
shifted due to non-circular motions (see \S~3.1). 
The wedge-shaped regions and the inner Galaxy are blanked out because 
no reliable distances to the \schi\ gas can be derived in these areas. 
Galactic longitudes ($l$) are marked at intervals of 15$\arcdeg$. 
The filled ellipse near the center indicates the central bar \citep{wegg2015}.
\label{fig:fig2}}
\end{figure}
\begin{figure}
\centering
\includegraphics[width=2.9in, trim=0 -3cm 0 -2cm]{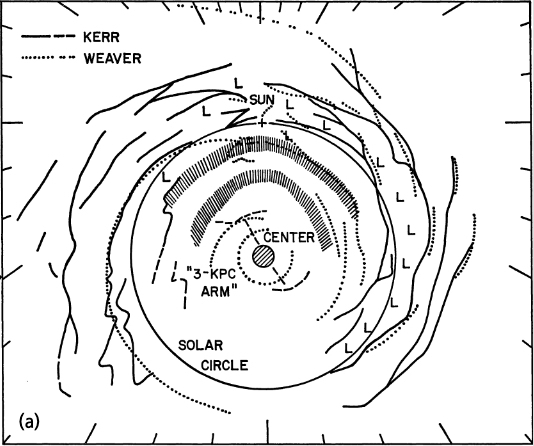}
\hspace*{1.8truecm} \includegraphics[width=3.4in, trim=0 0 0 0]{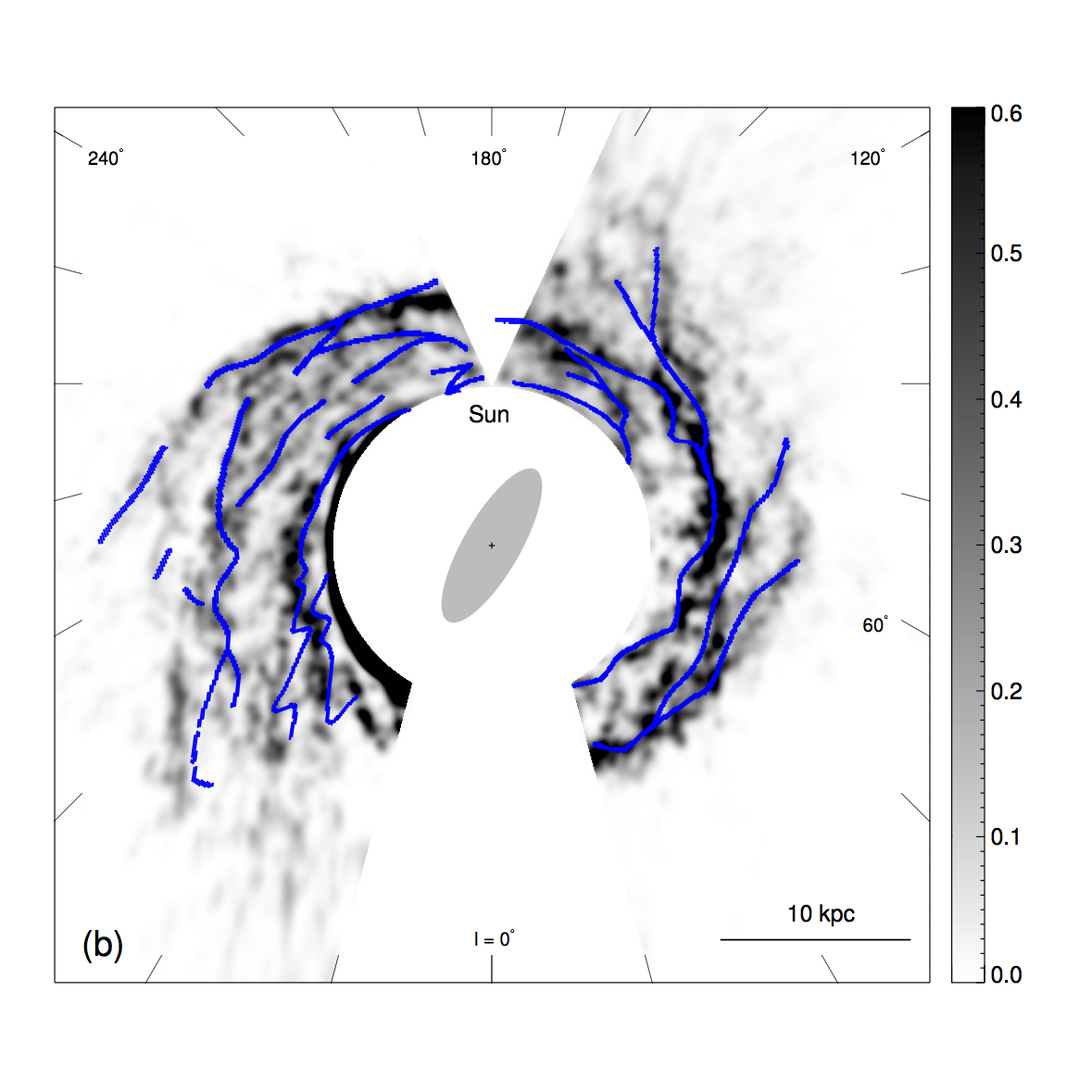}
\hspace*{-0.3truecm} \includegraphics[width=3.5in, trim=0.3cm -2.4cm 0 0]{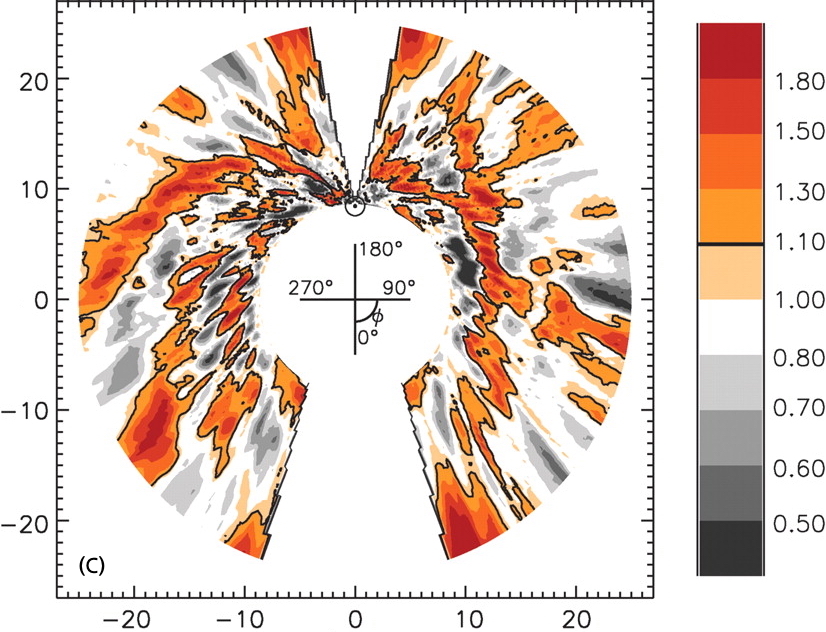}
\hspace*{0.3truecm} 
\includegraphics[width=3.4in, trim=0 0 0 0]{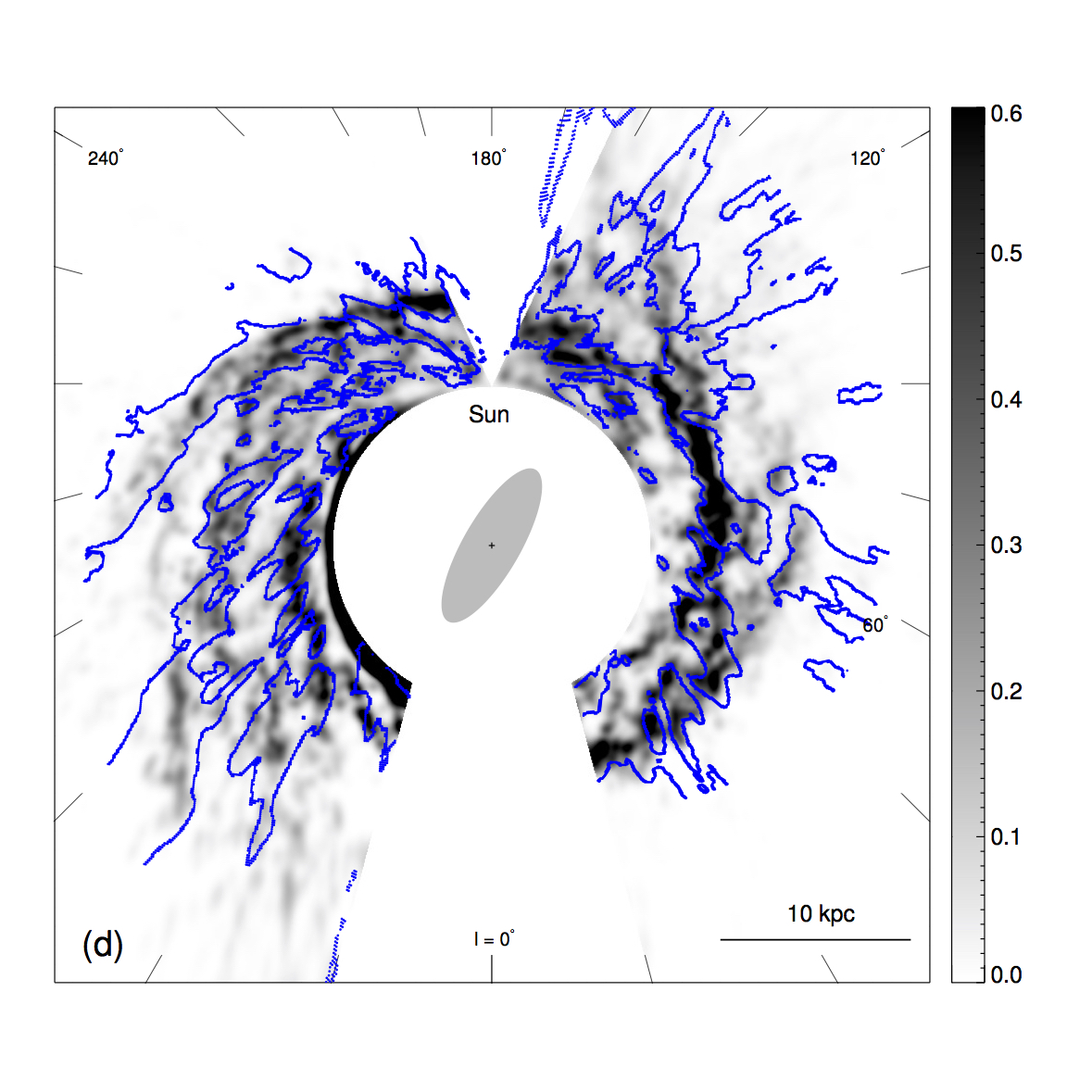}
\caption{
{(a)}
\linespread{0.6}\selectfont{}\small
Sketches of \schi\ ridges determined by \cite{kerr1969} and \cite{weaver1970} 
\citep[Figure 1 in][reproduced with permission 
 \textcopyright~ESO]{simonson1970}.
{(b)} Same as Figure \ref{fig:fig2} 
but with the Kerr's ridges overlaid. 
\cite{kerr1969} used the Schmidt rotation curve \citep{schmidt1965}, and 
his ridges are reprojected using the flat rotation curve  
that we adopted in this paper (see \S~2).
{(c)} Face-on map of  perturbed \schi\ surface density 
obtained by \cite{levine2006a} using an unsharp-masking technique 
\citep[Figure 1 in][reprinted with permission from AAAS]{levine2006a}. 
The color scale represents perturbed 
\schi\ surface density normalized by a local median value. 
The $x$ and $y$ scales are in kpc.   {(d)} Same as Figure~\ref{fig:fig2} but with the Levine's contours at 1.10 overlaid. 
\cite{levine2006a} used their own rotation curve (see \S~2),
and their contours are reprojected using the 
flat rotation curve that we adopted in this paper.
\label{fig:fig3}}
\end{figure}
\begin{figure} 
\centering
\includegraphics[width=4.8in, trim={0 3cm 0 2 cm}]{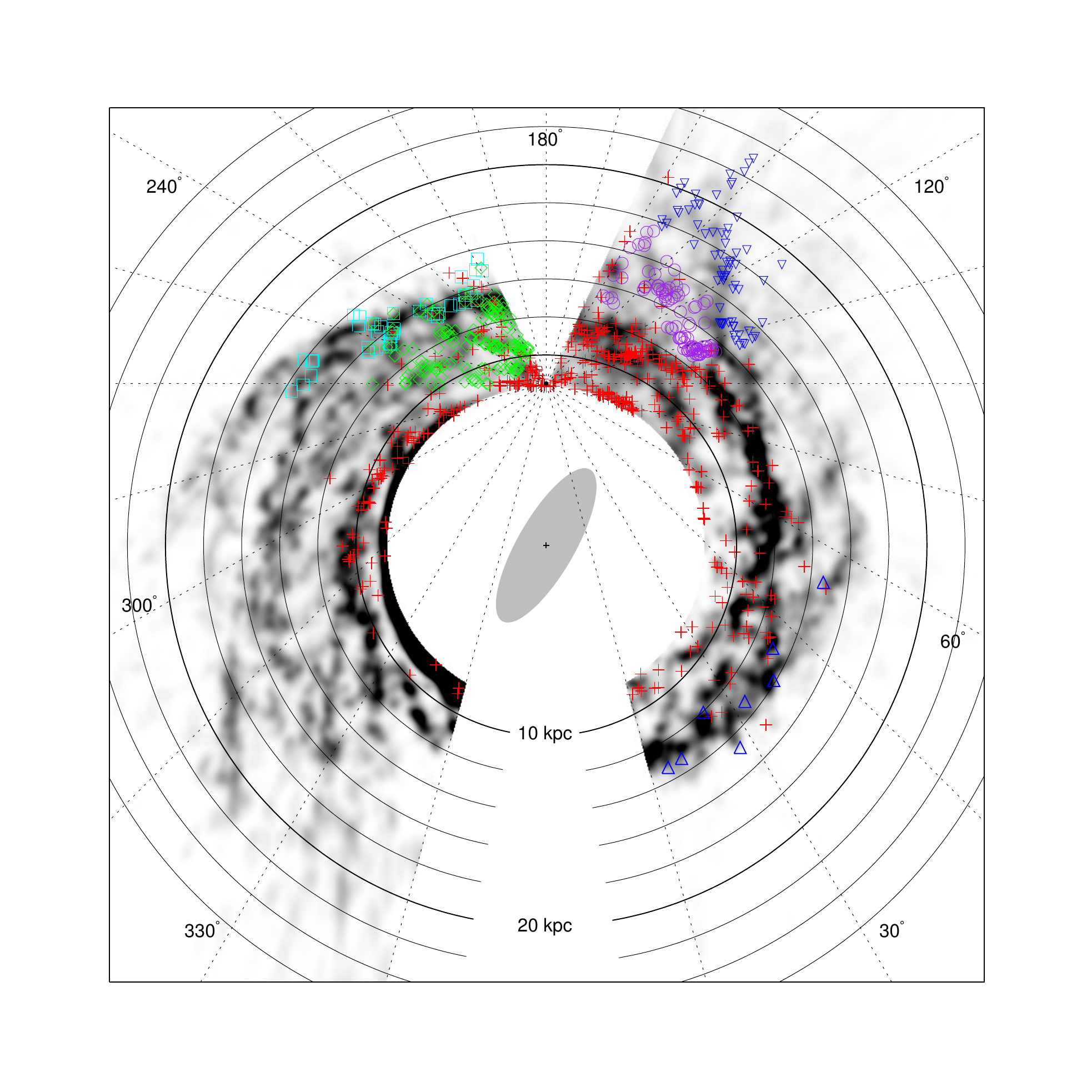}
\vspace{-1.0truecm}
\includegraphics[width=4.8in, trim={0 0 0 1 cm}]{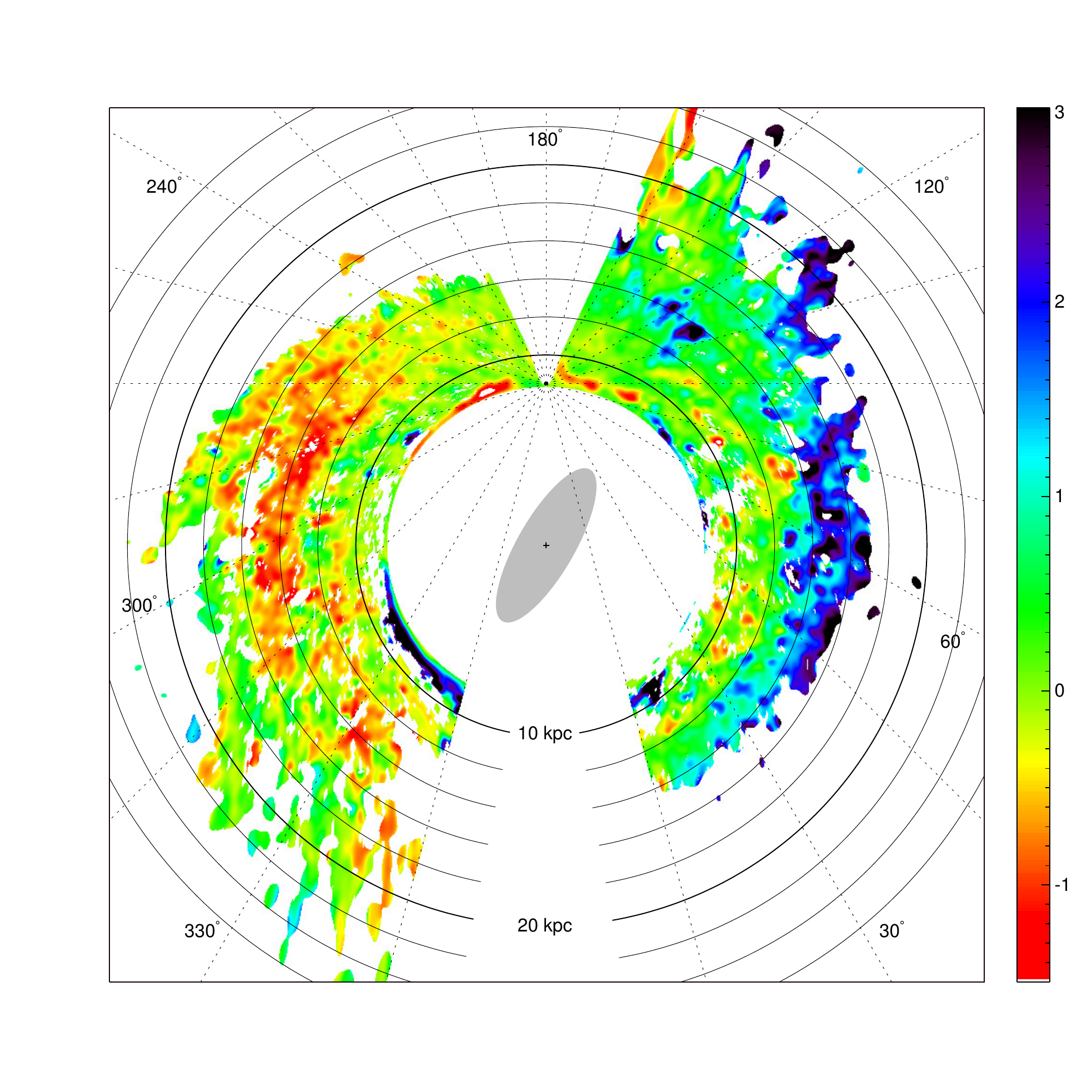}
\caption{\linespread{0.6}\selectfont{}\small
{\em Top:} \schi\ face-on map with \schii\ regions and CO clouds overlaid: red crosses $=$ {\em WISE} HII regions \citep{anderson2014}; 
other symbols $=$ 
CO clouds of \citet{may1997} (green diamonds),  
\citet{vazquez2008} (cyan squares),  
 \citet{dame2011} (blue triangles), 
\citet{sun2015} (blue upside-down triangles), and 
\citet{du2016} (purple circles). 
\schii\ regions with multiple velocity components are not drawn,
while only CO clouds with masses greater than $5\times 10^3$~\msol\ are shown for \citet{du2016}.
{\em Bottom:} Mass-weighted mean height, i.e., 
$\langle z \rangle \equiv \int z \rho(z)dz/\sigmahi$, of 
the \schi\ concentrations with $\sigmahi \ge 0.05$~$M_{\odot}$~pc$^{-2}$ in kpc. 
The color scale varies from $-1.5$~kpc to +3 kpc as in the color bar.  
\label{fig:fig4}} 
\end{figure}
\begin{figure}
\centering
\includegraphics[width=4.8in, trim={0 3cm 0 2 cm}]{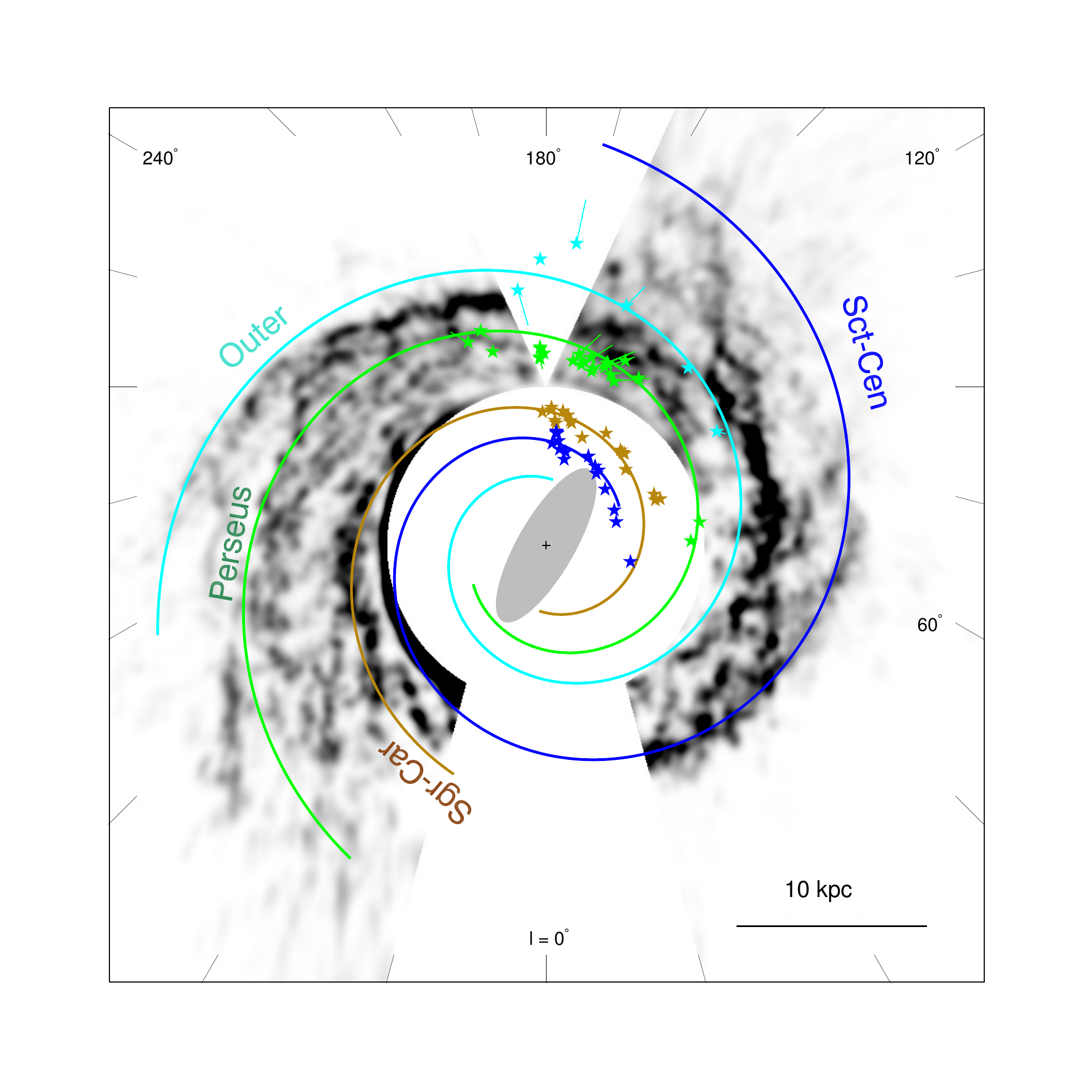}
\vspace{-1.0truecm}
\includegraphics[width=4.8in, trim={0 0 0 1 cm}]{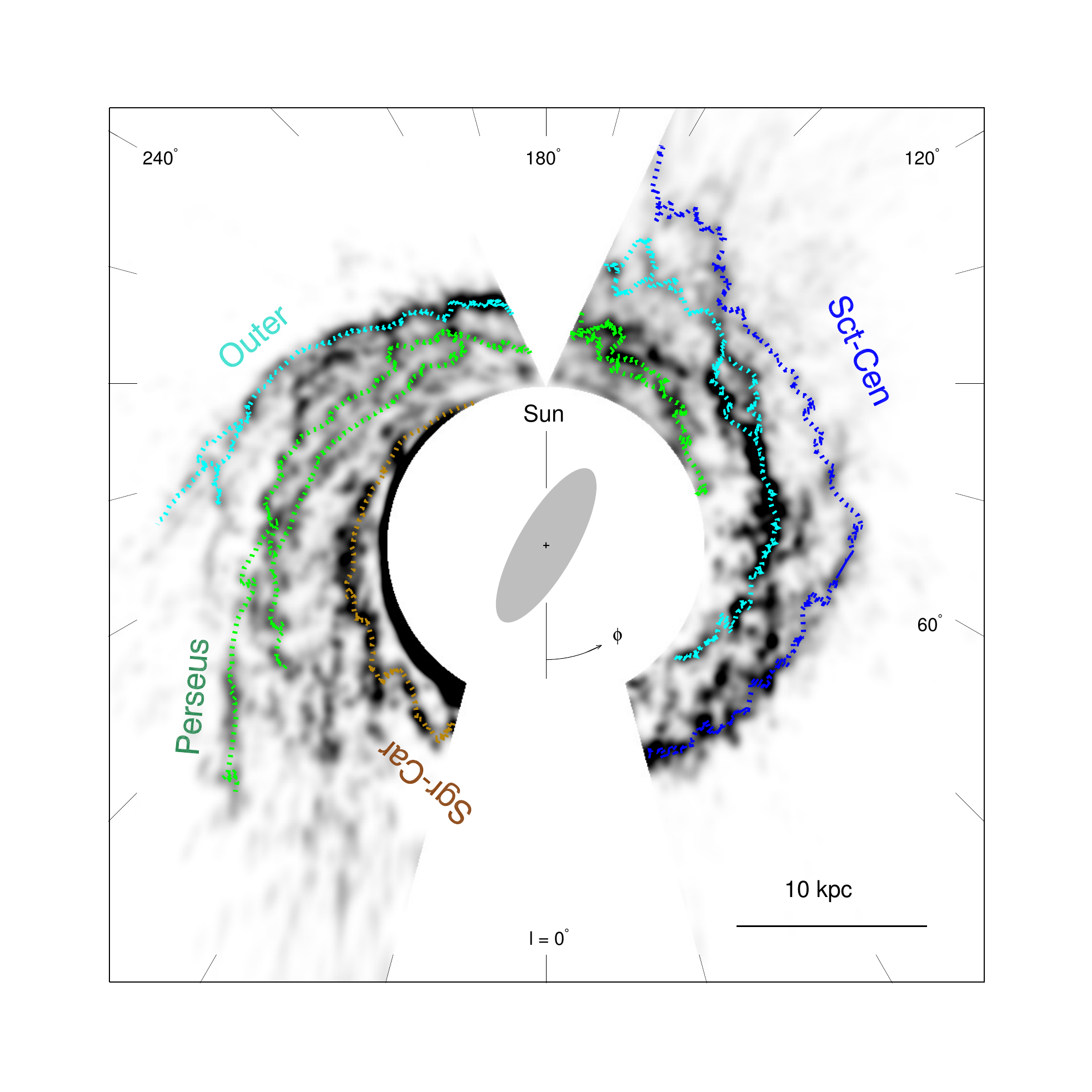}
\caption{
\linespread{0.6}\selectfont{}\small
{\em Top:} \schi\ face-on map with HMSFRs of parallax-determined distances \citep{reid2014} overlaid.  
The HMSFRs are color-coded according to their assigned spiral arms. 
The attached thin solid lines show the positional shift when their distances are 
determined from the LSR velocities.
The thick solid lines show a four-arm spiral model with a pitch angle 
of $12\fdg4$. The Perseus and Outer arms are obtained simply by rotating 
the other two arms by $180\arcdeg$. See \S~\ref{sec:res2} for more details.  
{\em Bottom:} \schi\ face-on map with spiral arm traces overlaid. 
The dotted line represents the \schi\ 
ridges of the major spiral arms determined in this 
work (see \S~\ref{sec:res2}).  
\label{fig:fig5}}
\end{figure}
\begin{figure}[t] 
\centering
\vspace*{5.0truecm}
\includegraphics[width=5.0in]{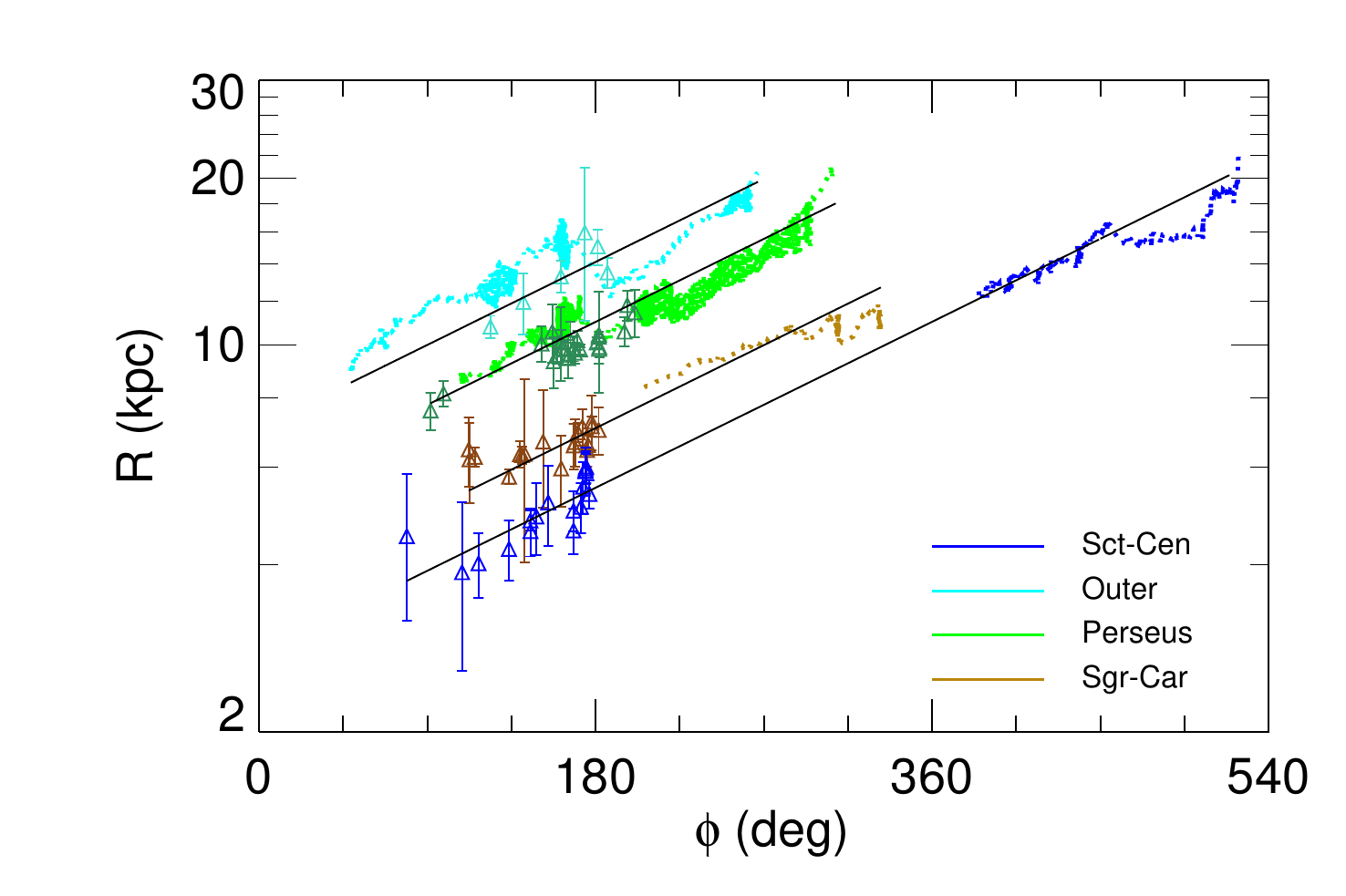}
\caption{ 
Galactocentric distance ($R$) versus azimuthal angle ($\phi$) of spiral tracers.
The dotted lines represent the \schi\ traces in Figure~\ref{fig:fig5}, while 
the empty triangles with error bars denote the locations of HMSFRs \citep{reid2014}.
The black straight lines correspond to the logarithmic spirals of pitch angle $12\fdg4$. 
The spirals crossing the Perseus and Outer arms are 
not the fits but the lines obtained by simply rotating the other two spirals by $180\arcdeg$ 
(see \S~\ref{sec:res2}). 
\label{fig:fig6}}
\end{figure}
\begin{figure}
\begin{center}
\includegraphics[width=6.0in]{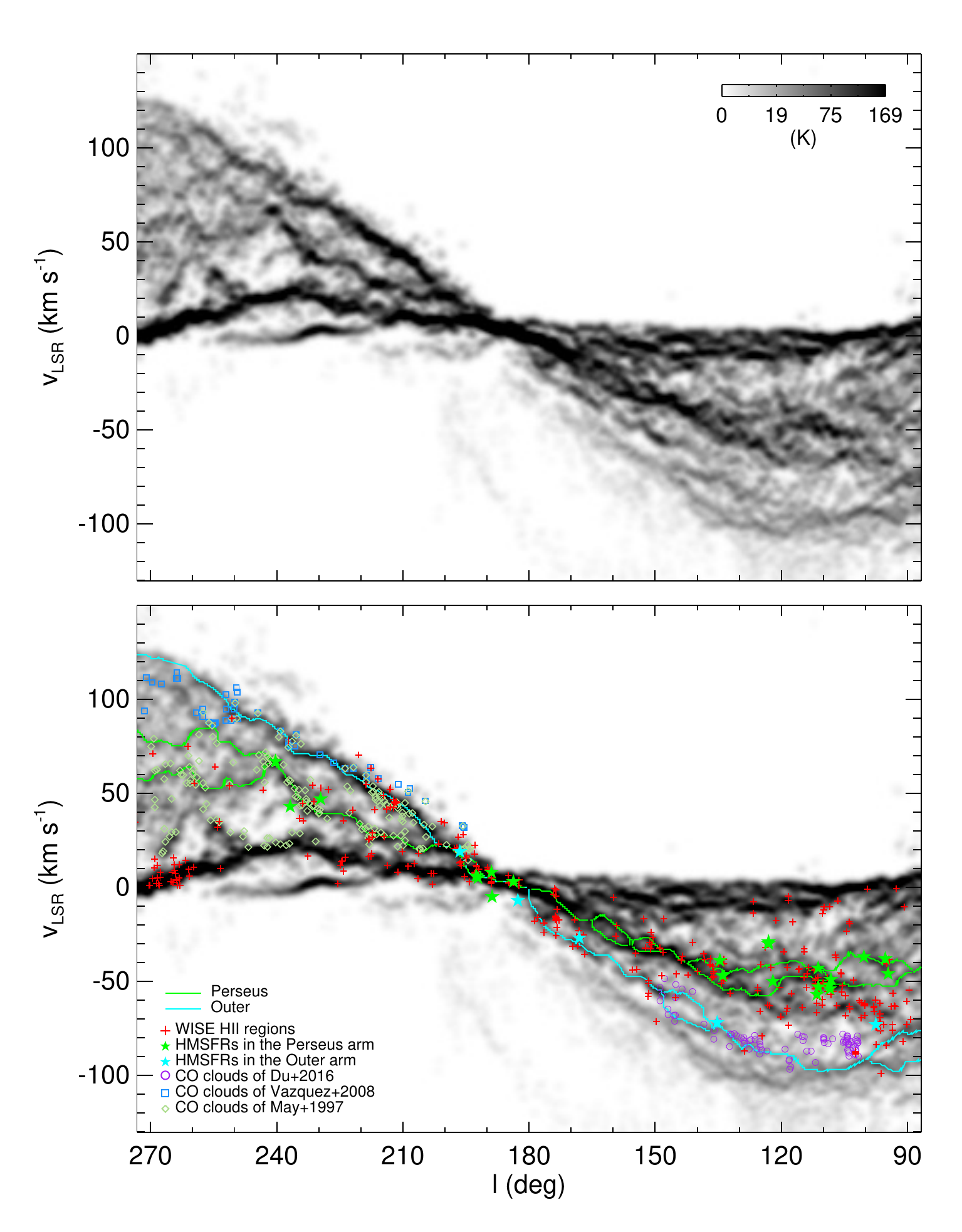}
\end{center}
\caption{ 
\linespread{0.6}\selectfont{}\small
{\it Top:} \schi\ $(l,\vlsr)$ diagram of the Galactic plane 
obtained by integrating local peaks 
from $b=-5\arcdeg$ to $+5\arcdeg$ 
(see \S~2 and Appendix for more details).
The grey scale represents the integrated intensity in K. 
{\it Bottom:} 
Same \schi\ $(l,\vlsr)$ diagram but with the ridges that we associate with  
the Perseus (green) and Outer (cyan) arms traced.
The CO clouds and \schii\ regions 
shown in Figure~\ref{fig:fig4} 
and the HMSFRs associated with 
the Perseus and Outer arms \citep[][see also Figure~\ref{fig:fig5}]{reid2016a} are 
also marked .  
\label{fig:fig7}
}
\end{figure}

Characterizing spiral structure in the outer Galaxy using strong
spiral tracers is limited, partly because high-resolution radio
observations with sufficient sensitivity are not yet available but
also because of the sparsity of those sources.  It had been suggested
that the intensity maxima of \schi\ 21-cm emission can trace spiral
structure, but this has not been explored with the sensitive all-sky
LAB data.  Here we verify that the face-on map of dense \schi\ 
concentrations obtained by projecting \schi\ 21-cm line peak intensities
is indeed a useful and efficient way to trace spiral structure in the
outer Galaxy. With the caveat that some arm structures 
may have been shifted and/or deformed due to non-circular motions, 
it is still a unique way to see the global spiral structure  
of the Milky Way far out beyond the solar circle. 
In the following, we summarize our main results:

\bigskip\noindent
(1) Our face-on map (Figure~\ref{fig:fig2}) 
strikingly  shows the complex \schi\ spiral structure of the outer Galaxy. 
The structure is characterized by 
several prominent long ($\simgt 20$~kpc) arcs  
that might be segments of major spiral arms and  
numerous ridges either 
connecting these spiral arm features or branching out from them.  
We confirm the identity of these \schi\ arc structures as spiral arms   
by comparing them 
to the distribution of other more conventional spiral tracers, e.g., 
\schii\ regions, CO clouds, and HMSFRs.
Some interarm features are prominent and 
have associated \schii\ regions.

\bigskip\noindent
(2) The global \schi\ spiral structure in the outer Galaxy is  
consistent with the four-arm spiral model with some deviations.
The Sgr-Car and Sct-Cen \schi\ arms have well-established \schi\ 
features, and they connect well 
to their HMSFR counterparts in the inner Galaxy by logarithmic spirals with a pitch angle of  $12\fdg$4. 
This pitch angle agrees with the global pitch angle of the spiral structure 
determined in previous studies.
For the Perseus arm, there is a faint but    
distinct \schi\ arm feature that can connect to the HMSFRs   
by a logarithmic spiral with the same pitch angle. 
For the Outer arm, there are bright \schi\ arc structures 
wobbling with respect to the logarithmic spiral arm. 
The \schi\ spiral features show complex structure 
near the ends;  the Perseus and Outer arms bifurcate, while  
the Sct-Cen arm bends inwards with negative pitch angle.  
Such variations of pitch angles along individual arms 
are similar to what we see in other galaxies.

\bigskip\noindent
(3) Our results provide an insight into identifying \schi\ features 
associated with individual spiral arms. 
In the southern Galaxy, there have been controversies in  
the identification of \schi\ arc structures corresponding to the 
Perseus and Outer arms. Our map clearly shows that there   
are two distinct \schi\ arc structures that can be identified as 
those two spiral arms.  According to our identification, in the southern Galaxy, 
the \schi\ Outer arm is much more prominent than 
the \schi\ Perseus arm which is relatively faint and indistinct. 

\bigskip\noindent
(4) In the Appendix, we show the locations of 
spiral arm traces in an $(l,\vlsr)$ diagram and compare them 
to the result of previous studies. 
Our traces are not very different from those in 
\cite{reid2016a} except for the Perseus/Outer arms in the third quadrant; Reid et al.
identified the \schi\ ridges at the 
most positive LSR velocities as the Perseus arm, while 
we consider that those ridges correspond to the Outer arm 
and that it is the \schi\ ridges at lower positive LSR velocities corresponding to the Perseus arm.
We also provide a catalog of prominent interarm structures in the Appendix. 


\acknowledgments

We thank Bob Benjamin for many valuable comments that helped preparing the manuscript. 
We also wish to thank Tom Dame, Mark Reid, and the anonymous referee 
for their comments and suggestions whch helped us improve the paper.
This work was supported by the National Research Foundation of Korea (NRF) grant 
funded by the Korea Government (MSIP) (No. 2012R1A4A1028713). 
The National Radio Astronomy Observatory is a facility of the
National Science Foundation operated under cooperative agreement by
Associated Universities, Inc.  T.V.W. is supported by the NSF through
the Grote Reber Fellowship Program administered by Associated
Universities, Inc./National Radio Astronomy Observatory, the
D.N. Batten Foundation Fellowship from the Jefferson Scholars
Foundation, the Mars Foundation Fellowship from the Achievement
Rewards for College Scientists Foundation, and the Virginia Space
Grant Consortium.




\software{IDL}

%

\newpage

\clearpage

\clearpage

\clearpage

\clearpage

\begin{deluxetable}{ccc}
\tablecaption{Parameters of Spiral Arms\label{table:tbl1}}
\tablewidth{30pt}

\tablecolumns{3} \tablehead{
\colhead{~~Arm~~} &\colhead{~~~~~Pitch angle~~~~~} & \colhead{~~$\phi_0$~~} \\ 
&   (deg)  &  (deg) }
\startdata            
Sgr-Car    &     $12\fdg4$      &      $223\arcdeg\pm 17\arcdeg$ \\ 
Perseus    &     $12\fdg4$       &      $108\arcdeg$  \\
Outer         &    $12\fdg4$       &         $43\arcdeg$ \\  
Sct-Cen     &    $12\fdg4\pm 1\fdg8$  &  $288\arcdeg\pm 47\arcdeg$ \\
\enddata
\tablecomments{These are the parameters of the spiral arms 
in Figure~\ref{fig:fig5}. The values without error bars represent  
fixed parameters.
The parameters of the Perseus and Outer arms are obtained simply by rotating 
the other two arms by $180\arcdeg$. See \S~\ref{sec:res2} for more details. 
}
\end{deluxetable}

\clearpage
\appendix

\section{Determination of Spiral Arm Traces}

\begin{figure}
\begin{center}
\includegraphics[width=5.2in]{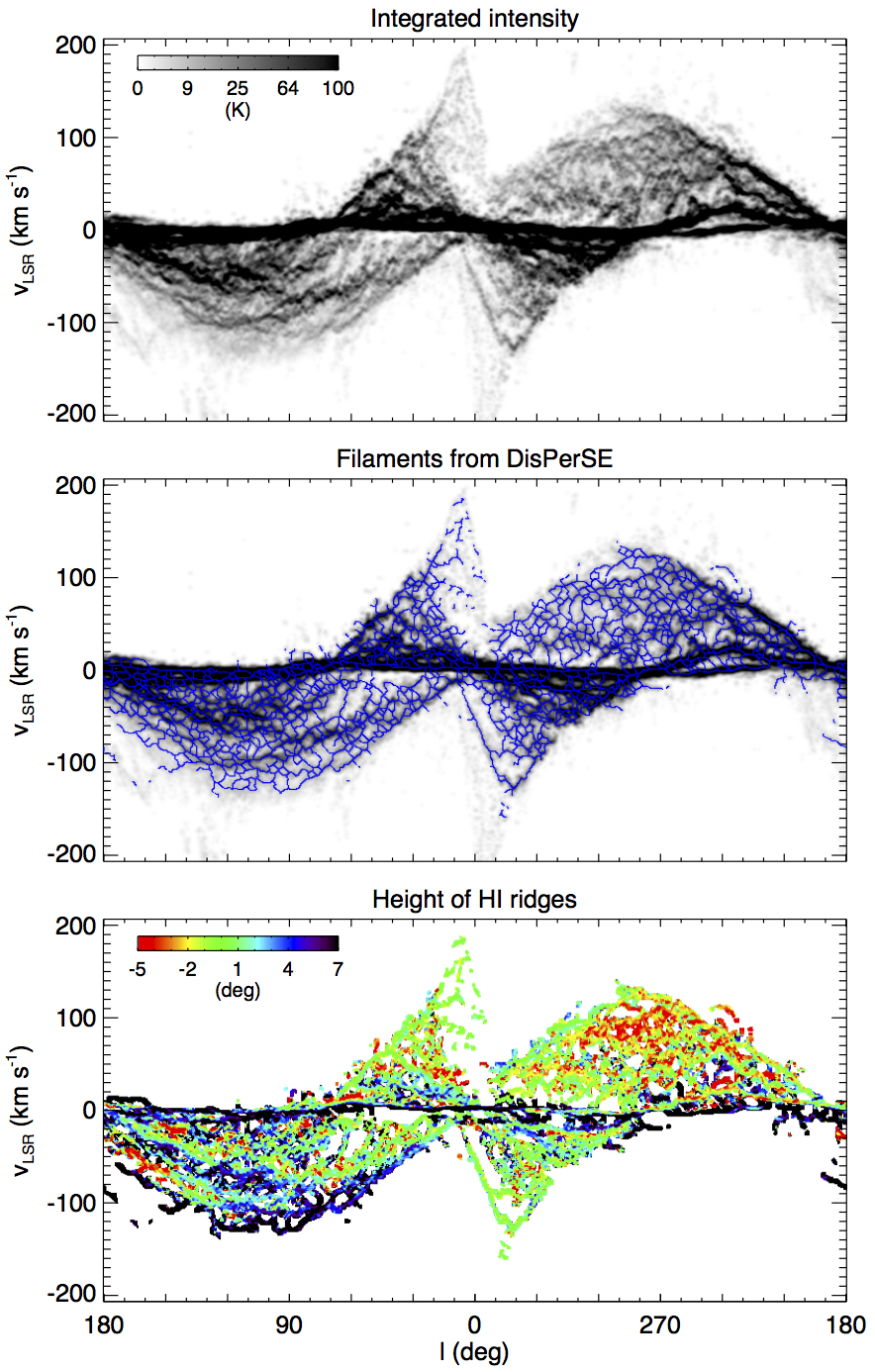}
\caption{
\linespread{0.6}\selectfont{}\small
{\it Top:} \schi\ $(l,\vlsr)$ diagram of the Galactic plane 
obtained by integrating local peaks from $b=-10\arcdeg$ to $+30\arcdeg$.
The grey scale represents the integrated intensity in K. 
{\it Middle:} Filaments   
found by the software \disperse\ overlaid on the ($l,\vlsr)$ diagram.
{\it Bottom:} 
Intensity-weighted mean height (degrees) of \schi\ ridges  
along the \disperse\ filaments. 
\label{fig:lvmap1}}
\end{center}
\end{figure}

The spiral arm traces shown in the face-on map in Figure~\ref{fig:fig5} (bottom frame) are
obtained in an $(l,\vlsr)$ diagram.  
The diagram was generated from the LAB \schi\ data 
by integrating the three dimensional cube
of local peak intensities $\tbmax(l,b,\vlsr)$ 
along the b axis from $b = -10\arcdeg$ to $+30\arcdeg$, 
which corresponds to 1.8~kpc and 5.8~kpc at a distance of 10 kpc, respectively.  
This should include all of the warped spiral features. 
The top frame in Figure~\ref{fig:lvmap1} shows the resulting $(l,\vlsr)$ diagram. 
We smoothed the map using a Gaussian kernel with FWHM of (3, 3) pixel 
or (150~pc, 3.06~\kms). 
Similar maps of the northern Galaxy were produced in earlier studies
\citep{kepner1970, weaver1970, davies1972, verschuur1973,weaver1974},
and more recently for the Sct-Cen and Perseus arm features \citep{reid2016a}. 
Our $(l,\vlsr)$ diagram, however, shows all spiral arm features in an `unbiased' single map.

We identified the ridges delineating the peak positions  
of integrated intensities using ``Discrete Persistent Structures Extractor (\disperse)''
\citep{sousbie2011}\footnote[4]{\url{http://thierry-sousbie.github.io/DisPerSE/}} which 
is open source software for the automatic 
identification of topological structures such as filaments and voids in 2D and 3D noisy data sets. 
The software first locates arcs connecting maxima/minima 
and saddle points, and calculates their `persistence' or the intensity difference between the two end points. 
Then the arcs with persistence greater than some threshold, e.g., 
the background intensity of that area, are considered as `persistent' topological features and 
they are connected to form filaments while the arcs with persistence less than the threshold are eliminated. 
In our case, after inspecting the persistence versus intensity 
plot of the arcs in the $(l,\vlsr)$ diagram, we adopted a threshold of 1~K 
to include faint features of the Sct-Cen arm 
but trimmed the arcs with peak intensities less than 2~K 
to remove strange and very faint features at high velocities,
e.g., below $-150$~\kms\ near $l\sim 130\arcdeg$.
The middle frame in Figure~\ref{fig:lvmap1} shows the 
filaments (blue contours) traced by \disperse\ in this way.

\begin{figure}
\begin{center}
\includegraphics[width=5.2in]{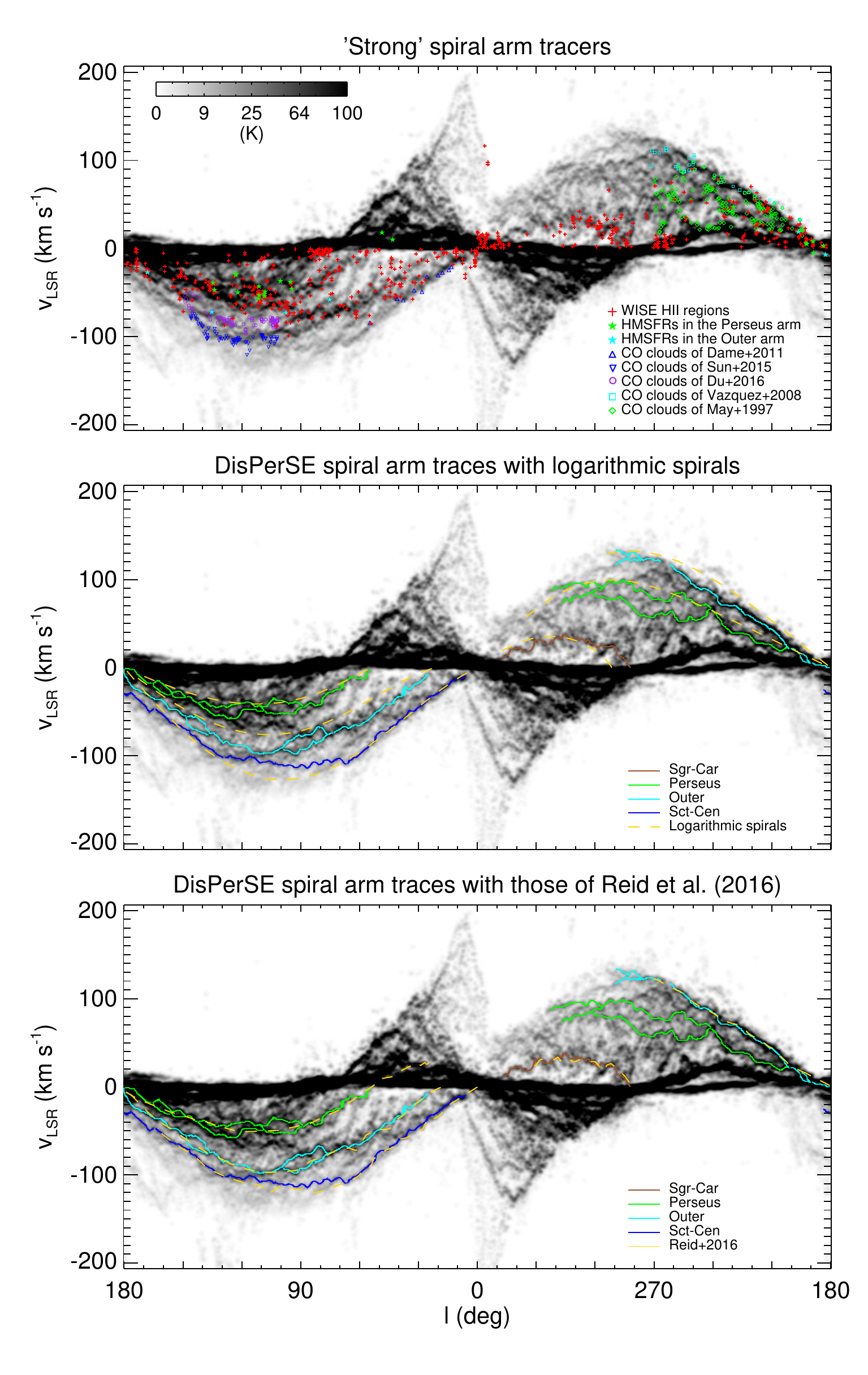}
\vspace*{-0.2truecm}
\caption{
\linespread{0.6}\selectfont{}\small
{\it Top:} Same \schi\ $(l,\vlsr)$ diagram as Figure~\ref{fig:lvmap1}
with other conventional spiral arm tracers marked for the outer Galaxy. 
The tracers are the same as shown in Figures~\ref{fig:fig4}--\ref{fig:fig5}.
The velocities of the HMSFRs are from Table~1 of \citet{reid2014}. 
{\it Middle:} The ridges selected as spiral arm traces (zigzagging solid lines)
are overlaid on the $(l,\vlsr)$ diagram.
The dashed yellow lines represent logarithmic spiral arms 
of pitch angle $12\fdg4$ obtained in the main text (Figure~\ref{fig:fig5}).
{\it Bottom:} Comparison of our spiral arm traces to those of \citet{reid2016a}. 
Note that the \schi\ ridges assigned to the Outer arm 
in the third quadrant were assigned to the Perseus arm by \citet{reid2016a}.
\label{fig:lvmap2}}
\end{center}
\end{figure}

\begin{figure}
\begin{center}
\epsscale{.9}
\plotone{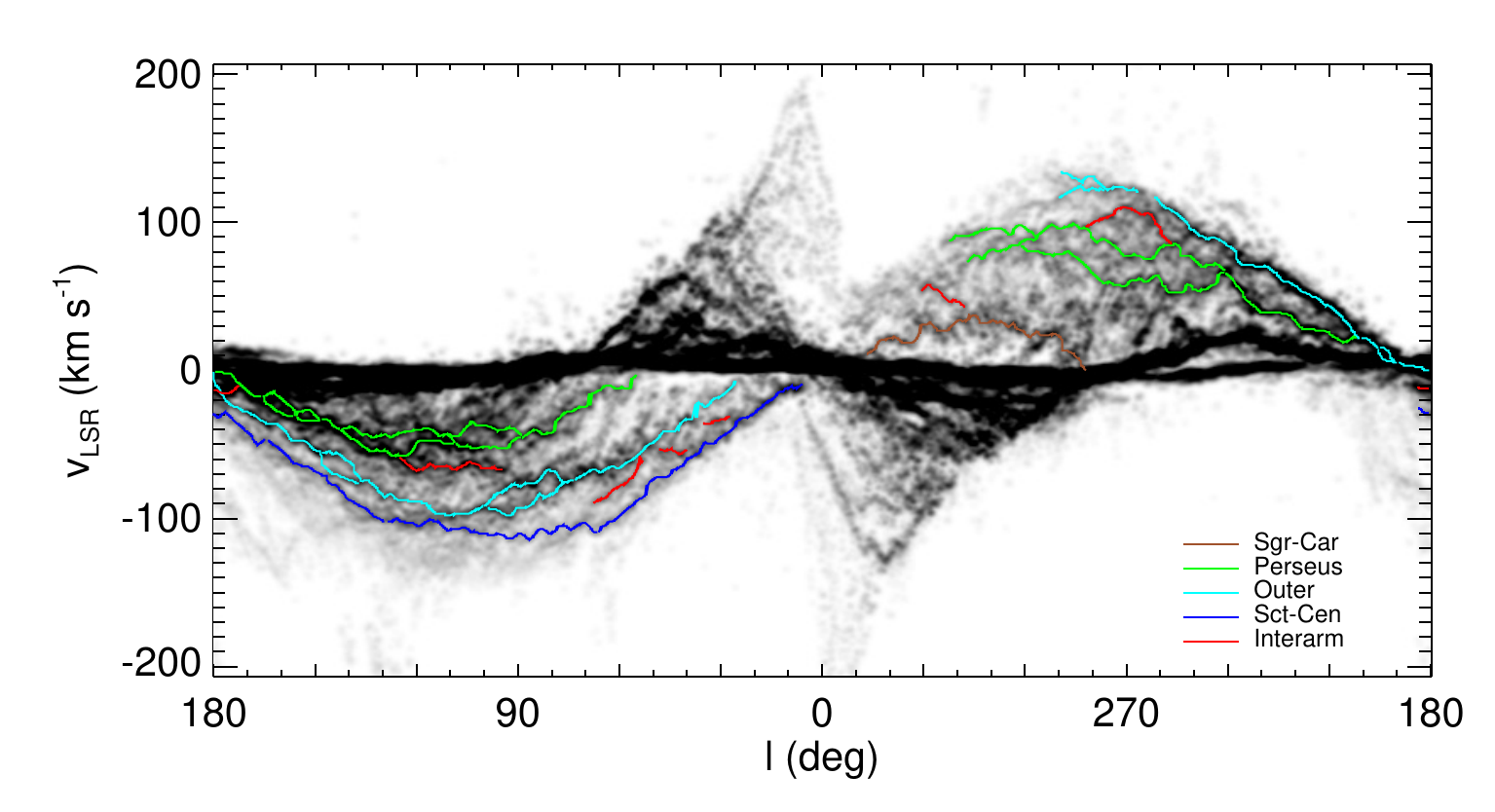}
\hspace*{1.2cm} \plotone{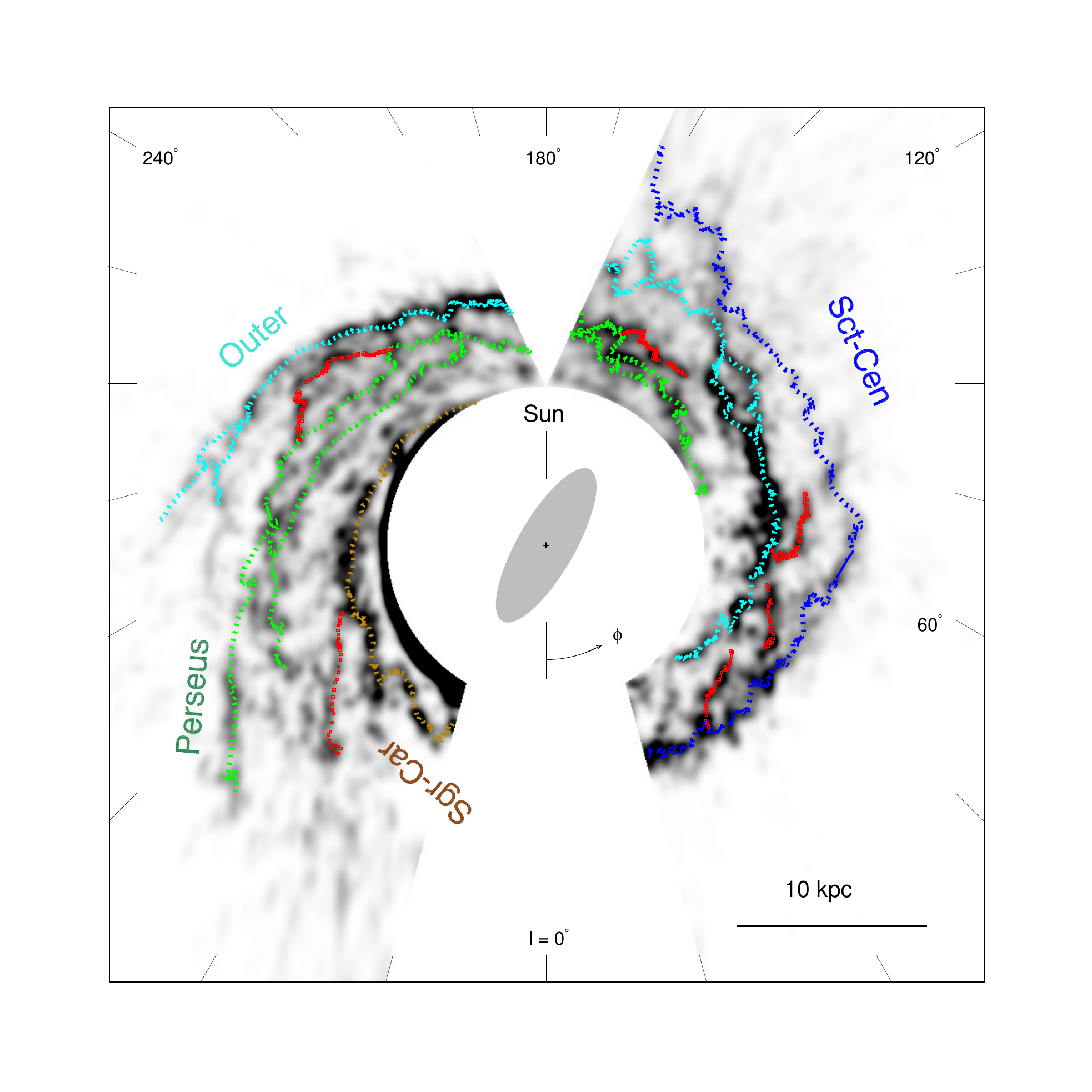}
\caption{The locations of the prominent interarm features   
in the $(l,\vlsr)$ diagram (top frame) and in 
the \schi\ face-on map (bottom frame).
The grey-scale face-on map is the same 
as shown in Figure~\ref{fig:fig5} but with 
the traces of the interarm features marked by red lines.
\label{fig:interarm}}
\end{center}
\end{figure}

The next step is to choose the ridges corresponding to the spiral arms.
As explained in the main text, the regularity of the 
observed \schi\ spiral structure cannot be described by either 
two- or three-arm models, and 
we identify the \schi\ ridges  corresponding to one of the arms 
in the four-arm spiral model, i.e., the model with the approximately 
equally spaced Sgr-Car, Perseus, Outer, and Sct-Cen arms
\citep{georgelin1976,wainscoat1992,churchwell2009,efremov2011,vallee2014,vallee2015}.
For the well-established arm features
or for the coherent long arc structures, this procedure is rather straightforward. 
But in the areas with complicated
structures or for the features of confusing identity, this procedure is
subjective.
The most challenging region is the second quadrant 
from $l=110\arcdeg$ to $180\arcdeg$, where the \schi\ ridges split and overlap.
Presumably, interarm features as well as 
non-circular motions due to streaming motions along spiral arms 
or some large-scale motions with peculiar velocities  
might be responsible for such complex structures.
The bottom image of Figure~\ref{fig:lvmap1}, which shows the 
mean galactic latitude of \schi\ ridges ($\bar b$), 
is helpful because $\bar b$ should be continuous along an arm and 
should be close to the latitude of the Galactic midplane. 
The presence of other conventional spiral tracers, e.g., 
young stars, star-forming regions, \schii\ regions, and molecular clouds are also helpful 
(Figure~\ref{fig:lvmap2} top frame). 
When there are more than one possible branches, we adopted them both.
Figure~\ref{fig:lvmap2} shows the adopted ridges (solid lines). 
Adjoining ridges are grouped into segments for convenience, 
and their start and end locations are listed in Table~\ref{tab:sparm} together with their mean galactic latitudes. 
There are also bright \schi\ ridges 
located between the adopted arm ridges, e.g., 
$(l,\vlsr)=(40\arcdeg,-54~{\rm km~s}^{-1})$ to $(48\arcdeg,  -54~{\rm km~s}^{-1})$ 
between the Outer and Sct-Cen arms, and  
$(l,\vlsr)=(95\arcdeg,-67~{\rm km~s}^{-1})$ to $(125\arcdeg, -59~{\rm km~s}^{-1})$, and  
$(l,\vlsr)=(257\arcdeg,+86~{\rm km~s}^{-1})$ to $(282\arcdeg, +97~{\rm km~s}^{-1})$ 
between the Perseus and Outer arms.
We consider them interarm features.
Table~\ref{tab:interarm} lists prominent interarm features and
Figure~\ref{fig:interarm} shows their locations.
Some of them have associated \schii\ regions indicating 
on-going star formation (see Figure~\ref{fig:fig2}).

In the Figure~\ref{fig:lvmap2} middle and bottom images, 
we overlay the logarithmic spirals with pitch angles of $12\fdg4$ 
(see main text) and the traces of \citet{reid2016a}, respectively. 
We will compare our result to that of \citet{reid2016a} for individual arms below,
but in general our traces are not very different from their traces except 
for the Perseus arm in the third quadrant.
\citet{reid2016a} identified the \schi\ ridges at the most positive velocities 
as the Perseus arm, but we consider that those ridges correspond to the Outer arm and 
that it is the \schi\ ridges at lower positive velocities corresponding to the Perseus arm.
In the following, we summarize the characteristics of the traces of 
individual spiral arms:

\bigskip\noindent
(1) Sgr-Car arm

The Sgr-Car arm is traced from $l=282\arcdeg$ to $l=347\arcdeg$, and this 
is a well known spiral arm feature from early studies \citep{kerr1969, weaver1970}. 
The segment at $l\simlt 330\arcdeg$ is almost identical to that of \citet{reid2016a} 
determined from the CO distribution, while the segment at $l\simgt330\arcdeg$ 
is seen only in \schi. There is a systematic discrepancy 
between the \schi\ trace and the logarithmic spiral: 
the trace is located at a velocity higher (lower) than the model at longitudes 
less than (greater than) $310\arcdeg$.
The deviation is less than 10~\kms\ mostly, but at 
$l\simlt 290\arcdeg$ it becomes as large as 25~\kms. 
This deviation, however, does not yield a significant offset in real space
(see Figure \ref{fig:fig5}). 

\bigskip\noindent
(2) Perseus arm

The Perseus arm is traced from $l=55\arcdeg$ to $l=323\arcdeg$.
At several longitude intervals, multiple branches are allocated, e.g., 
at $l=89\arcdeg$--$142\arcdeg$ and $241\arcdeg$--$323\arcdeg$.
The Perseus arm \schi\ features are well known from early \schi\ studies \citep{kerr1969, weaver1970},
although the features are complex and their allocations are still controversial.
Our traces in the first and second quadrants are not very different from that of \citet{reid2016a} 
while in the third quadrant, our trace is located 
at $\sim15$--$30$~\kms\ lower velocities.
The Perseus arm segment of \citet{reid2016a} in the third quadrant corresponds to the Outer arm in our case.  
The segments in the third and fourth quadrants were recently reidentified by 
\citet{mgriffiths2004} as a separate spiral arm, i.e., the `Distant'  spiral arm. 
The pronounced long ridges with CO clouds, \schii\ regions, and HMSFRs supports 
our identification (see \S~\ref{sec:res3}).
At $l=181\arcdeg$--$201\arcdeg$, \schi\ traces of the Perseus and Outer arms
overlap each other because of the velocity crowding.
Many {\it WISE} \schii\ regions are detected along the Perseus arm, but 
at $l\simgt 270\arcdeg$,
essentially none has been detected.
Note that {\it WISE} \schii\ regions that considered here are those with measured velocities,
but the original {\it WISE} catalog lists many \schii\ regions with no available velocity information.
So, the absence of \schii\ regions should be checked by further observations.

The Perseus arm traces in the first and second quadrants are complex and located at 
systematically higher negative velocities than the logarithmic spiral arm model.
It has long been known that the Perseus arm in the second quadrant 
shows non-circular motion of typically $\sim 10-15$~\kms\   
\citep[e.g.,][]{miller1968, humphreys1976, xu2006, russeil2007,reid2014}. 
Some of the \citeauthor{reid2014}'s HMSFRs at $l=100\arcdeg$--$135\arcdeg$ show
larger peculiar velocities ($\simgt 20$~\kms).
The lower branch at $l=241\arcdeg$--$323\arcdeg$ also has a large ($\simlt 30$~\kms) 
velocity departure.  

\bigskip\noindent
(3) Outer arm

The Outer arm is traced from $l=26\arcdeg$ to $l=290\arcdeg$.
At some longitude intervals, multiple branches are allocated, e.g., at 
$l=74\arcdeg$--$100\arcdeg$ and $137\arcdeg$--$149\arcdeg$.
The Outer arm \schi\ ridges had been also identified 
in early \schi\ studies \citep{kerr1969, weaver1970,davies1972}.
On the other hand, \citet{reid2016a} identified these segments as the Perseus arm.
As we mention above, however, we identify these segments as being
part of the Outer arm since there is a separate, well-defined,
segment corresponding to the Perseus arm (see \S~\ref{sec:res3}).
Almost a continuous distribution of CO clouds along the \schi\ ridges has been detected  
in $l=100\arcdeg$--$150\arcdeg$ \citep{du2016}.
Between $l=100\arcdeg$ and $120\arcdeg$, there are 
also CO clouds not associated with \schi\ ridges 
at somewhat lower ($\sim 20$~\kms) negative velocities.
There are many \schii\ regions along the arm  at $l\simlt 100\arcdeg$ and 
also along the bright \schi\ ridges beyond.

The \schi\ trace of the Outer arm is not well fit by a logarithmic spiral.
There is a systematic offset between the arm traces and any logarithmic arm model:   
the \schi\ trace has a higher negative velocity 
than the model in the first and second quadrants while it has a lower positive velocity  
than the model in the third and fourth quadrants. 
For our logarithmic spiral model described in the main text,
the deviation in the third and fourth quadrants is   
$\simlt 20$~\kms, while it is considerably larger ($\simlt 30$~\kms) in the first and second quadrants.  

\bigskip\noindent
(4) Sct-Cen arm

The Sct-Cen arm is traced from $l=6\arcdeg$ to $l=184\arcdeg$.
The bright segment between $l=6\arcdeg$ and 
$80\arcdeg$ had been identified in early \schi\ studies \citep{kerr1969,weaver1970,weaver1974}.
Our trace agrees with that of \citet{reid2016a} except at $l=66\arcdeg$--$120\arcdeg$ 
where ours is located at slightly ($\simlt 10$~\kms) lower negative velocities. 
Our trace also extends further, i.e., to 
$l=184\arcdeg$ compared to $l=165\arcdeg$ of \citet{reid2016a}. 
CO clouds associated with the segments at $l=10\arcdeg$--$70\arcdeg$ 
and $l=100\arcdeg$ and $150\arcdeg$ 
have been detected \citep{dame2011,sun2015}. 
There are several \schii\ regions along the former segment. 

The linear \schi\ segment 
at $l\simlt 66\arcdeg$ (Sct-Cen\_s1) is fit very well by a logarithmic spiral, while 
at $l=66$--$130\arcdeg$, the 
trace significantly ($\simlt 20$~\kms) deviates from 
the logarithmic spiral.
There is a faint \schi\ feature that can be matched to the 
logarithmic spiral arm in this longitude range, 
but its $\bar b$ increases continuously beyond $l=90\arcdeg$ 
and therefore is not likely a main spiral feature. 
(The warping of the Galactic midplane peaks at $l\sim 90\arcdeg$ and decreases 
as we move away along azimuth \citep{levine2006b}.)
The segment at $l=165\arcdeg$--$184\arcdeg$ (Sct-Cen\_s4) is well defined, but 
it is located at considerably higher negative velocities than the model. 
In the 4th quadrant, relevant \schi\ features 
may be there but it is not clearly seen in the $(l,\vlsr)$ diagram 
shown in the paper due to the contamination by  
the strong emission from the  local \schi\ gas in the solar neighborhood. 

%

\clearpage
\begin{deluxetable}{p{1.cm}lc rrc rrc c}
\tabletypesize{\scriptsize}
\tablewidth{0pt}
\tablecaption{\small Segments of Spiral Arm Traces\label{tab:sparm}}
\tablehead{
 & && \multicolumn{2}{c}{Start Location} && 
\multicolumn{2}{c}{End Location} && \\
\cline{4-5} \cline{7-8}
{\#\dotfill} & {Name} && \colhead{$\ell$} & \colhead{$v_{\rm LSR}$}  &&
    \colhead{$\ell$} & \colhead{$v_{\rm LSR}$} && \colhead{Mean $\bar b$} \\
 & && \colhead{($^\circ$)} & \colhead{(\kms)}  &&
    \colhead{($^\circ$)} & \colhead{(\kms)} && \colhead{($^\circ$)}   
}
\startdata
 1\dotfill & Srg-Car\_s1                      && 282.0 &$+$0.0   && 346.5 &$+$11.3  &&$-$0.4 (1.9)  \\
 2\dotfill & Perseus\_s1                      &&  55.0 &$-$4.1   && 88.5  &$-$45.4  &&$+$0.6 (1.7)  \\
 3\dotfill & Perseus\_s2\_1                   &&  89.0 &$-$45.4  && 108.5 &$-$49.5  &&$+$0.4 (1.2)  \\
 4\dotfill & Perseus\_s2\_2                   &&  89.0 &$-$44.3  && 111.0 &$-$42.3  &&$+$1.1 (2.2)  \\
 5\dotfill & Perseus\_s3\_1                   && 109.0 &$-$49.5  && 142.5 &$-$40.2  &&$+$1.7 (1.8)  \\
 6\dotfill & Perseus\_s3\_2                   && 108.5 &$-$48.4  && 142.0 &$-$40.2  &&$-$0.2 (1.7)  \\
 7\dotfill & Perseus\_s4                      && 143.0 &$-$40.2  && 148.0 &$-$33.0  &&$+$0.7 (1.1)  \\
 8\dotfill & Perseus\_s5\_1                   && 148.5 &$-$34.0  && 155.5 &$-$32.0  &&$+$2.1 (0.7)  \\
 9\dotfill & Perseus\_s5\_2                   && 148.5 &$-$33.0  && 156.0 &$-$27.8  &&$-$0.4 (0.9)  \\
10\dotfill & Perseus\_s6\_1                   && 156.5 &$-$30.9  && 165.5 &$-$17.5  &&$+$1.5 (1.4)  \\
11\dotfill & Perseus\_s6\_2                   && 156.0 &$-$30.9  && 164.5 &$-$17.5  &&$+$0.1 (1.4)  \\
12\dotfill & Perseus\_s7                      && 166.0 &$-$17.5  && 179.5 &$-$1.0   &&$+$3.0 (1.3)  \\
13\dotfill & Perseus\_s8     && 180.5 &$+$0.0   && 189.5 &$+$5.2   &&$+$1.8 (1.2)  \\
14\dotfill & Perseus\_s9\_1  && 190.5 &$+$6.2   && 195.0 &$+$15.5  &&$+$2.8 (5.3)  \\
15\dotfill & Perseus\_s9\_2  && 190.0 &$+$5.2   && 195.5 &$+$11.3  &&$+$0.6 (2.7)  \\
16\dotfill & Perseus\_s10    && 195.5 &$+$15.5  && 200.5 &$+$21.6  &&$-$0.5 (0.6)  \\
17\dotfill & Perseus\_s11                     && 202.0 &$+$21.6  && 240.5 &$+$64.9  &&$-$0.5 (1.3)  \\
18\dotfill & Perseus\_s12\_1                  && 241.0 &$+$67.0  && 301.5 &$+$85.5  &&$-$2.1 (1.3)  \\
19\dotfill & Perseus\_s12\_2                  && 241.0 &$+$66.0  && 301.0 &$+$83.5  &&$-$1.9 (1.4)  \\
20\dotfill & Perseus\_s13\_1                  && 301.5 &$+$84.5  && 317.0 &$+$73.2  &&$-$2.3 (1.1)  \\
21\dotfill & Perseus\_s13\_2                  && 297.0 &$+$89.7  && 322.5 &$+$87.6  &&$-$0.9 (1.1)  \\
22\dotfill & Outer\_s1                        &&  25.5 &$-$7.2   &&  73.0 &$-$74.2  &&$+$0.3 (1.1)  \\
23\dotfill & Outer\_s2\_1                     &&  73.5 &$-$74.2  && 100.0 &$-$91.7  &&$+$2.3 (2.3)  \\
24\dotfill & Outer\_s2\_2                     &&  76.0 &$-$76.3  &&  99.0 &$-$92.8  &&$+$3.0 (1.3)  \\
25\dotfill & Outer\_s3                        && 100.5 &$-$92.8  && 136.0 &$-$75.2  &&$+$2.0 (0.8)  \\
26\dotfill & Outer\_s4\_1                     && 136.5 &$-$75.2  && 148.0 &$-$55.7  &&$+$2.4 (1.9)  \\
27\dotfill & Outer\_s4\_2                     && 137.0 &$-$76.3  && 148.5 &$-$56.7  &&$+$0.9 (0.7)  \\
28\dotfill & Outer\_s5                        && 148.5 &$-$55.7  && 180.0 &$-$6.2   &&$+$2.1 (3.2)  \\
29\dotfill & Outer\_s6       && 180.5 &$+$0.0   && 189.5 &$+$5.2   &&$+$1.8 (1.2)  \\
30\dotfill & Outer\_s7\_1    && 190.5 &$+$6.2   && 195.0 &$+$15.5  &&$+$2.8 (5.3)  \\
31\dotfill & Outer\_s7\_2    && 190.0 &$+$5.2   && 195.5 &$+$11.3  &&$+$0.6 (2.7)  \\
32\dotfill & Outer\_s8       && 195.5 &$+$15.5  && 200.5 &$+$21.6  &&$-$0.5 (0.6)  \\
33\dotfill & Outer\_s9                        && 201.0 &$+$21.6  && 276.0 &$+$121.6 &&$-$1.6 (1.2)  \\
34\dotfill & Outer\_s10\_1                    && 276.5 &$+$121.6 && 290.0 &$+$116.5 &&$-$1.8 (0.7)  \\
35\dotfill & Outer\_s10\_2                    && 276.0 &$+$121.6 && 289.5 &$+$134.0 &&$-$1.3 (1.1)  \\
36\dotfill & Sct-Cen\_s1                      &&   6.0 &$-$10.3  &&  66.0 &$-$109.2 &&$+$3.2 (1.6)  \\
37\dotfill & Sct-Cen\_s2                     &&  66.5 &$-$109.2 && 128.5 &$-$102.0 &&$+$4.7 (2.2)  \\
38\dotfill & Sct-Cen\_s3                      && 129.5 &$-$102.0 && 164.0 &$-$47.4  &&$-$1.5 (2.4)  \\
39\dotfill & Sct-Cen\_s4                      && 165.0 &$-$47.4  && 183.5 &$-$25.8  &&$-$2.1 (4.2)  \\
\enddata                         
\tablecomments{
This table lists the segments assigned to the spiral arms where 
Name=name of arm segment, Start and End locations = ($l,\vlsr)$ values of the 
start and end locations of the segements, and 
Mean $\bar b=$ mean latitude of the segment with standard deviation in parenthesis.
For multiple branchs, additional numbers are attached to the name, i.e., 
s2\_1 and s2\_2, etc.}
\end{deluxetable}
\clearpage

\begin{deluxetable}{p{1.cm}lc rrc rrc cc}
\tablecolumns{11}
\tabletypesize{\scriptsize}
\tablewidth{0pt}
\tablecaption{\small Traces of Prominent Interarm Features\label{tab:interarm}}
\tablehead{
 & && \multicolumn{2}{c}{Start Location} && \multicolumn{2}{c}{End Location} && & \\
\cline{4-5} \cline{7-8}
{\#\dotfill} & {Name} && \colhead{$\ell$} & \colhead{$v_{\rm LSR}$}  &&
    \colhead{$\ell$} & \colhead{$v_{\rm LSR}$} && \colhead{Mean $\bar b$} & {Arm?} \\
 & && \colhead{($^\circ$)} & \colhead{(\kms)}  &&
    \colhead{($^\circ$)} & \colhead{(\kms)} && \colhead{($^\circ$)} &
}
\startdata
1\dotfill & Interarm\_s1 &&  25.5 &  $-$36.1 &&  35.0 &  $-$35.0 && $+$0.8 (1.2) & Sct-Cen        \\ 
2\dotfill & Interarm\_s2 &&  40.0 &  $-$53.6 &&  48.0 &  $-$53.6 && $+$1.5 (0.4) & Outer, Sct-Cen \\ 
3\dotfill & Interarm\_s3 &&  53.0 &  $-$60.8 &&  67.5 &  $-$89.7 && $+$3.4 (1.5) & Outer          \\ 
4\dotfill & Interarm\_s4 &&  94.5 &  $-$67.0 && 125.0 &  $-$58.7 && $+$3.1 (1.5) & Perseus        \\ 
5\dotfill & Interarm\_s5 && 172.5 &  $-$10.3 && 184.0 &  $-$11.3 && $+$5.9 (4.5) & Perseus        \\ 
6\dotfill & Interarm\_s6 && 256.5 &  $+$85.5 && 282.0 &  $+$96.9 && $-$2.7 (1.2) & Perseus, Otuer \\ 
7\dotfill & Interarm\_s7 && 318.0 &  $+$42.3 && 330.5 &  $+$53.6 && $-$1.3 (1.1) & Sgr-Car        \\ 
\enddata                         
\tablecomments{The parameters in columns 
3--7 are the same as those in Table~\ref{tab:sparm}.  
The last column lists the spiral arm(s) that might be relevant  
(see Figure \ref{fig:interarm}).}
\end{deluxetable}
\clearpage

\end{document}